\documentclass[aip,jcp,preprint] {revtex4-1} 

\usepackage{graphicx}
\usepackage{subfigure}
\usepackage{xcolor}
\usepackage{enumerate}
\usepackage{braket}
\usepackage{amsmath, bm}
\usepackage{stackengine}
\usepackage{adjustbox}
\usepackage{rotating}
\usepackage{array}

\usepackage{color, soul}
\definecolor{darkblue}{rgb}{0,0,0.5}
\setulcolor{darkblue}
\begin{document}

\title{
Benchmark Computations of Nearly Degenerate Singlet and Triplet states of N-heterocyclic Chromophores : II. Density-based Methods}

\author{Shamik Chanda$^{\dagger}$}
\affiliation{Department of Chemical Sciences\\
Indian Institute of Science Education and Research (IISER) Kolkata \\  
Nadia, Mohanpur-741246, WB, India}
\author{Subhasish Saha$^{\dagger}$}
\affiliation{Department of Chemical Sciences\\
Indian Institute of Science Education and Research (IISER) Kolkata \\  
Nadia, Mohanpur-741246, WB, India}
\affiliation{MLR Institute of Technology, \\ Dundigal, Hyderabad-500043, Telangana, India}
\author{Sangita Sen}
\altaffiliation{Email: sangita.sen@iiserkol.ac.in }
\affiliation{Department of Chemical Sciences\\
Indian Institute of Science Education and Research (IISER) Kolkata \\  
Nadia, Mohanpur-741246, WB, India}

\begin{abstract}
 In this paper we demonstrate the performance of several density-based methods in predicting the inversion of S$_1$ and T$_1$ states of a few N-heterocyclic fused ring molecules (popularly known as INVEST molecules) with an eye to identify a well performing but cheap preliminary screening method. Both conventional LR-TDDFT and $\Delta$SCF methods (namely MOM, SGM, ROKS) are considered for excited state computations using exchange-correlation (XC) functionals from different rungs of the Jacob's ladder. A well-justified systematism is observed in the performance of the functionals when compared against FICMRCISD and/or EOM-CCSD, with the most important feature being the capture of spin-polarization in presence of correlation. A set of functionals with the least mean absolute error (MAE) is proposed for both the approaches, LR-TDDFT and $\Delta$SCF, which can be cheaper alternatives for computations on synthesizable larger derivatives of the templates studied here. We have based our findings on extensive studies of three cyclazine-based molecular templates, with additional studies on a set of six related templates. Previous benchmark studies for subsets of the functionals were conducted against the DLPNO-STEOM-CCSD, which resulted in an inadequate evaluation due to deficiencies in the benchmark theory. The role of exact-exchange, spin-contamination and spin-polarization in the context of DFT comes to the forefront in our studies and supports the numerical evaluation of XC functionals for these applications. Suitable connections are drawn to two and three state exciton models which identify the minimal physics governing the interactions in these molecules. 


$^\dagger$Contributed equally to this work.

\end{abstract}
\maketitle   

\section{Introduction}
Detecting candidates for organic light-emitting diodes (OLEDs) through the screening of potential chromophores is a cost-effective alternative to synthesis and experimental studies. OLEDs which are based on thermally activated delayed fluorescence (TADF), require the singlet–triplet splitting to be of the order of thermal energy. If the spin–orbit coupling (SOC) strength is sufficient, reverse inter-system crossing (RISC), i.e. the switch from lower-lying T$_1$ (dark state) to higher S$_1$ (bright state) is feasible leading to an enhancement in quantum yield. A new generation of proposed light-emitting molecules (some nitrogen substituted triangulenes like cyclazine, heptazine, etc.) for OLEDs has raised considerable research interest due to their exceptional feature - a negative singlet–triplet (ST) gap which violates the Hund’s rule of maximum multiplicity. RISC is preferable to ISC, for these systems, since it requires no thermal activation and can potentially provide a 100\% quantum yield. Nonetheless, it is crucial to bear in mind that vibronic coupling is also a part of the RISC process, and it gets smaller as the energy gap widens \cite{dinkelbach2021}. Therefore, the ultimate purpose of theoretically screening OLED candidate molecules for big inverted gaps may not be served. Furthermore, the experimental values for the few possible synthesised chemicals show essentially degenerate positive/negative ST gaps \cite{jacs80,jacs86,nature2022,ricci2021}. Modelling this near degeneracy has proven to be challenging  for theoretical methods. Some substituted derivatives of heptazine which had been theoretically predicted to have large inverted STG have been recently synthesized and are found to exhibit nearly degenerate although inverted STGs\cite{ehrmaier2019,nature2022}. 
The efficiency and temperature dependence of TADF are not solely dependent on the STG\cite{ehrmaier2019}, as recent theoretical calculations of rate constants have shown\cite{dinkelbach2021}, but almost degenerate S$_1$ and T$_1$ states, have been shown to be ideal for effective RISC in OLEDs\cite{dinkelbach2021}.

\begin{figure}
    \centering
    \includegraphics[width=1.0\linewidth]{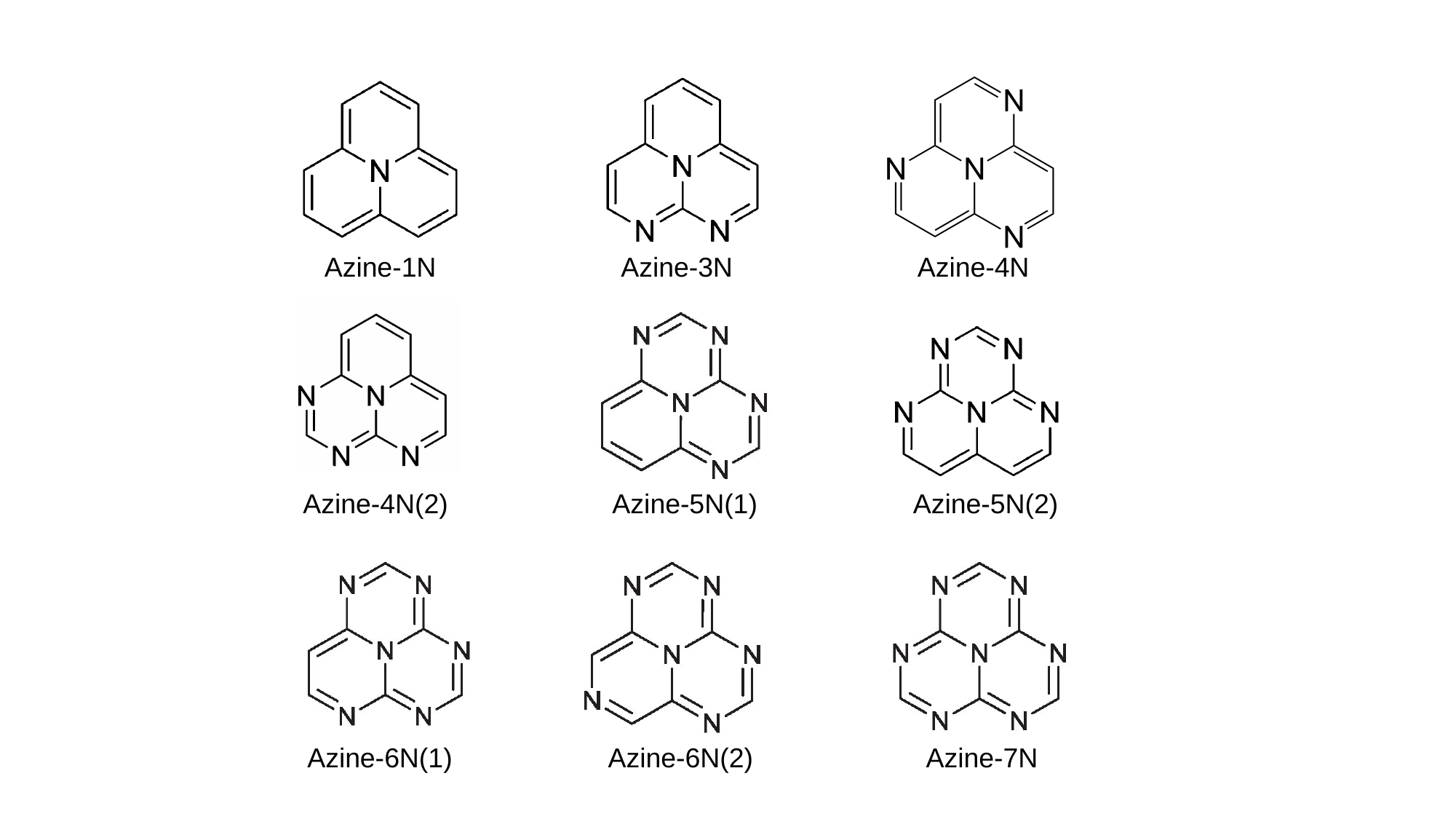}
    \caption{Chemical Structure of molecules}
    \label{molecule}
\end{figure}

T$_1$ is usually below the S$_1$ state with $\Delta E_{ST} > 0$ due to the exchange interactions that stabilise the T$_1$ state (Hund's Law\cite{hund1925}). Nevertheless, greater amount of electron correlation in S$_1$ relative to T$_1$ reduces the ST gap in these N-substituted fused ring molecules (popularly known as INVEST molecules, such as, cyclazine (azine-1N), heptazine (azine-7N) and their derivatives. See Fig.~1), resulting in degenerate or inverted S$_1$-T$_1$ gaps which have been verified by experiments\cite{jacs80,jacs86,nature2022}. 
Popular linear response excited state methods such as Configuration Interaction Singles (CIS)\cite{pople2003,foresman1992}, Random Phase Approximation (RPA)\cite{bouman1983,bouman1989} or Time-Dependent Density Functional Theory (TD-DFT)\cite{casida1995} which are formulated for treating primarily singly excited states\cite{dreuw2005} almost always give this gap as positive\cite{ghosh2022,desilva2019b}. 
Recently, in this context, azine-1N, azine-4N and azine-7N (heptazine) (see Fig.~\ref{molecule}) based molecules have been studied widely in the electronic structure community\cite{ghosh2022,desilva2019b,ricci2021,li2022,police2021,nature2022,domcke2021,dreuw2023,rodrigo2021,ehrmaier2019,pios2021,tuckova2022,loos2023,bedogni2024,sanchogarchia2022,dinkelbach2021,loos2023}. The effect of ST inversion is effectively captured, to varying degrees, by the inclusion of higher order correlations through several single and multi-reference theories, as we have addressed in detail in our prior work\cite{chanda2023nature}.
It has been observed that the domain-based local pair natural orbital (DLPNO) similarity transformed EOM-CCSD (STEOM-CCSD)\cite{nooijen1997}, which is computationally less expensive, can also invert the gaps but unfortunately the STEOM approximation distorts the physics and further acceleration with the DLPNO approximation makes it worse, leading to much larger negative values than the parent EOM-CCSD\cite{bhattacharyya2021,ghosh2022}.
While most of the earlier investigations have concentrated on which methods predict a larger inversion, we have recently discovered that higher levels of theory predict near degeneracy of S$_1$ and T$_1$ with -0.2 $<$ $\Delta$E$_{ST}$ $<$ 0.01 eV, which is more in line with experimental results\cite{jacs80,jacs86,nature2022}. Our previous work\cite{chanda2023nature} reveals that a proper balance between static and dynamic correlation, leads us from the large positive $\Delta$E$_{ST}$ of uncorrelated wavefunction theories to the large negative $\Delta$E$_{ST}$ of CIS(D) and RPA(D) and ultimately to the near degeneracy of EOM-CCSD/CCSDT, CASSCF, NEVPT2, and FIC-MRCISD. EOM-CCSD is quite good in most cases, but deviates from the FIC-MRCISD and/or experimental results in a few cases which appears to stem from the absence of triples as EOM-CCSDT as well as multi-reference correlated theories readily rectify the number. CASSCF and NEVPT2 are also somewhat inconsistent across various molecules. Among the wave function based methods, FIC-MRCISD is the most accurate method considered by us, so far. A complete active space (CAS) consist of 12 electrons and 9 orbitals (6 occupied and 3 virtual orbitals) are considered for CASSCF computations in our benchmark study and further dynamic correlations through FICMRCISD are included on top of this. 
The multi-reference treatment obviates the need for higher excitations in the dynamic correlation part.
Similar conclusions are drawn in Ref\cite{dreuw2023}, that the STGs of heptazine, cycl[3.3.3]azine, grow less negative the closer we examine them, or, to put it another way, the more precisely we attempt to compute them with any improvement made to the basis set or electron correlation level in the theoretical description of the molecules. It demonstrates that higher levels of theory capture greater electron-hole entanglement but reduce the inverted STG in magnitude\cite{dreuw2023}. Triples and the higher excitations are often responsible for dynamic spin-polarization for a fixed set of orbitals.  

Improvement of XC functionals typically involve inclusion of more \%HF excahnge. However, it has been observed that, regardless of the degree of non-local Hartree–Fock exchange (HFX), the linear-response time-dependent density functional theory (LR-TDDFT) with standard exchange-correlation (XC) kernels consistently produce positive STGs for heptazine\cite{bhattacharyya2021,ricci2021,tuckova2022}. 
Negative STGs can only be achieved by applying the spin-flip variation of TDDFT or by using double-hybrid long-range-corrected xc-functionals\cite{police2021,tuckova2022}. Heptazine's STG has been determined to be nearly zero at the DFT/MRCI level, (--0.01 eV\cite{dinkelbach2021}). In addition to the traditional TDDFT, it is now feasible to obtain the S$_1$ state within the single determinant framework of KS-DFT by applying a special orbital optimization method to converge the SCF process to the excited state configuration. These methods are commonly known as $\Delta$SCF methods with the well-known ones are being perturbative SCF\cite{baruah2009,baruah2012}, maximum overlap method (MOM)\cite{gilbert2008}, square-gradient minimization method (SGM)\cite{hait2020,hait2021}, restricted open-shell KS (ROKS)\cite{hait2016}, excited state DFT (e-DFT)\cite{edft2008}, constrained DFT (CDFT)\cite{cdft2006}, etc. In this work, we report that the $\Delta$SCF  methods, which use either the Maximum Overlap Method (MOM)\cite{gilbert2008} or the Square Gradient Minimization (SGM)\cite{hait2020} techniques, can cheaply and efficiently capture ST inversion and can be used as a preliminary screening tool.


The fundamental goals of the present work are two fold. First, to suggest through systematic benchmarking, the DFT-based approaches and XC-functionals best suited for screening the so called INVEST molecules. Secondly, we wish to understand the minimum physics that need to be captured for a satisfactory description of the INVEST molecules to aid in more detailed studies of promising candidates. In this paper, the S$_1$ and T$_1$ excitation energies are calculated through both standard LR-TDDFT and $\Delta$SCF frameworks (MOM and SGM) for several classes of exchange-correlation (XC) functionals (such as Local Density Approximation (LDA), Generalized Gradient Approximation (GGA),
meta-GGA, Range-Separated Hybrid Generalized Gradient Approximation (RSH-GGA), RSH-meta-GGA, Double Hybrid Generalized Gradient Approximation (DH-GGA) etc.). These are benchmarked against the FICMRCI numbers reported in our previous work\cite{chanda2023nature} for the three molecular templates (namely azine-1N,4N and 7N) in Fig.~1. A larger set of molecules along with the previous three are also considered as a secondary test set for verification and they are benchmarked against the EOM-CCSD numbers. A set of functionals with the least mean absolute error (MAE) against the benchmark numbers are suggested on the basis of these analysis. The role of \%HF exchange for several range-separated hybrid functionals are also analyzed. In the $\Delta$SCF methods, \%HF exchange has been found to be linked to the spin-contamination of the variational excited state. Spin-contamination was found to be crucial for proper energetics of these molecules in our previous study\cite{chanda2023nature}. Removing spin-contamination through a spin-pure ROKS or RO-SGM framework is also studied to know the role of spin-contamination in these systems. The HOMO-LUMO exchange-integral (K$_{if}$) between the unpaired electrons are also evaluated for several XC functionals as a potential indicator of INVEST behaviour.

Thus our objectives are to comprehend the fundamental mechanism of the inverted singlet–triplet gap in a density functional framework  and thereby identify low-cost access to excited states of INVEST molecules. 
A brief overview of the theoretical framework of TDDFT and $\Delta$SCF methods are documented in Sec.~II. 
The computational and technical details are given in Sec~III. In Sec~IV.A, a detailed analysis of the results obtained from TDDFT for several XC functionals are given including the double-hybrid functionals. Then, in Sec~IV.B, a detailed analysis of $\Delta$SCF results for several XC functionals are given. In the next section, Sec~IV.C, the role of \%HF exchange and spin-contamination along with the role of HOMO-LUMO exchange coupling (K$_{if}$) are analyzed to understand the underlying physics of the INVEST systems. Finally, the necessary findings along with the future outlook is provided in Sec~V.

\section{Theoretical Frameworks}
Excitation energies in the LR-TDDFT regime can be estimated using the non-Hermitian eigenvalue problem given by, 
\begin{equation}
\begin{bmatrix}
A & B \\
B^{\ast} & A^{\ast} \\
\end{bmatrix}
\begin{bmatrix}
X \\
Y\\
\end{bmatrix}
=\Omega_{TDDFT}
\begin{bmatrix}
1 &  0 \\
0 & -1 \\
\end{bmatrix}
\begin{bmatrix}
X \\
Y \\
\end{bmatrix}
\end{equation}

where $\Omega_{TDDFT}$ is the excitation energy matrix. X and Y are the respective single-particle excitation and de-excitation amplitudes. The A and B matrices for a general hybrid exchange-correlation (XC) functional are given as\cite{casida2012,dreuw2005,HERBERT2023},
\begin{equation}
    A_{ia,jb} = \delta_{ij}\delta_{ab} (\varepsilon_i - \varepsilon_a) + (ia|jb) - C_{HF} (ij|ab) + (1-C_{HF})(ia|f_{xc}|jb)
\end{equation}
\begin{equation}
    B_{ia,jb} =(ia|bj) - C_{HF} (ib|aj) + (1-C_{HF})(ia|\hat{f_{xc}}|jb)
\end{equation}

where occupied (unoccupied) orbitals are denoted as i,j(a,b). $C_{HF}$ is the fraction of Hartree-Fock (HF) exchange in the exchange-correlation functional. In Mulliken notation $(\textit{i}a\vert \textit{j}b)$ is the exchange integral $\iint \phi_i^\ast(\textbf{r})\phi_a(\textbf{r}) 1/|r-r'| \phi_j^\ast(\textbf{r}')\phi_b(\textbf{r}')d\textbf{r}d\textbf{r}'$, and the Coulomb integral is denoted as $(\textit{i}\textit{j}\vert ab)$. If $B=0$ then we get the corresponding Tamn-Dancoff approximations (TDA). The last term in these equations has a DFT origin, where the XC kernel, denoted by $\hat{f_{xc}}$, is defined as the second derivative of the XC functional $E_{XC}$ with respect to the electronic density $\rho$. 
\begin{equation}
    \hat{f_{xc}}(\textbf{r},\textbf{r}')= \frac{\partial^2E_{XC}[\rho]}{\partial\rho(\textbf{r})\partial\rho(\textbf{r}')}
\end{equation}

$C_{HF}$ in the above equations (2) and (3), is equal to 1 in the case of RPA or time-dependent HF. 
Furthermore if one employs TDA, we get the configuration interaction singles (CIS) formalism. The singlet-triplet energy difference is of primary importance to us, and has the form\cite{desilva2019a}, 
\begin{equation}
    \Delta E_{ST}= 2(ia|jb)+2(ia|\hat{f_{xc}}|ib)
\end{equation}

This result (Eq. (5)) was previously emphasized in refs\cite{desilva2019b,Moral2015} for a ``two-state model". The first term (two-electron integrals) in Eq.(5) is positive definite, 
while the adiabatic kernel $f_{xc}(r,r',\omega)$ is negative definite\cite{beckereview} leading to a reduction of the ST gap and ideally to inversion. But, for commonly known functionals this gap is always $>0$ as sufficient correlation is not captured. It is well-known that TD-DFT fails to describe double excitations if a frequency-independent kernel is used (ie. under the adiabatic approximation)\cite{elliott2011}. On the other hand, double hybrid functionals\cite{grimme2007} introduce double excitations through a post-SCF MP2 correction and can give an inverted ST gap\cite{police2021,sanchogarchia2022,ghosh2022}. 
The general expression of exchange-correlation energy in the case of double-hybrid functional
is given by 
\begin{equation}
    E_{XC}^{DH}=(1-a_{X})E_X^{DFA}+a_XE_X^{HF}+b 
  E_C^{DFA}+a_CE_C^{MP2}
\end{equation}
where $a_{X}$ is the scaling parameter for the HF exchange, b and $a_C$ scale the contributions of the density functional and perturbative correlations, respectively. $E_X^{DFA}$ and $E_X^{HF}$ stand for the local density functional and HF exchange, respectively. $E_C^{DFA}$ and $E_C^{MP2}$ are the local density functional correlation and nonlocal correlation respectively. $a_C$ is set to zero for usual hybrid functionals. Vertical excitation energies at the double-hybrid level are obtained in a two-step process; first, solve the TDDFT (Eq. 1) eigenvalue problem only for the hybrid part of the double hybrid density functionals (DHDF), i.e., all terms in Eq. 6 excluding the perturbative portion (last term). This results in a hybrid-DFT quality vertical excitation energy ($\Omega_{TDDFT}$), which is subsequently corrected by merging second-order perturbative correlation contributions ($\Delta_{(D)}$).

\begin{equation}
    \Omega_{TDDFT(D)}=\Omega_{TDDFT}+a_{C}\Delta_{(D)}
\end{equation}

\begin{figure}[h]
\centering

  \centering
	\includegraphics[width=\textwidth]{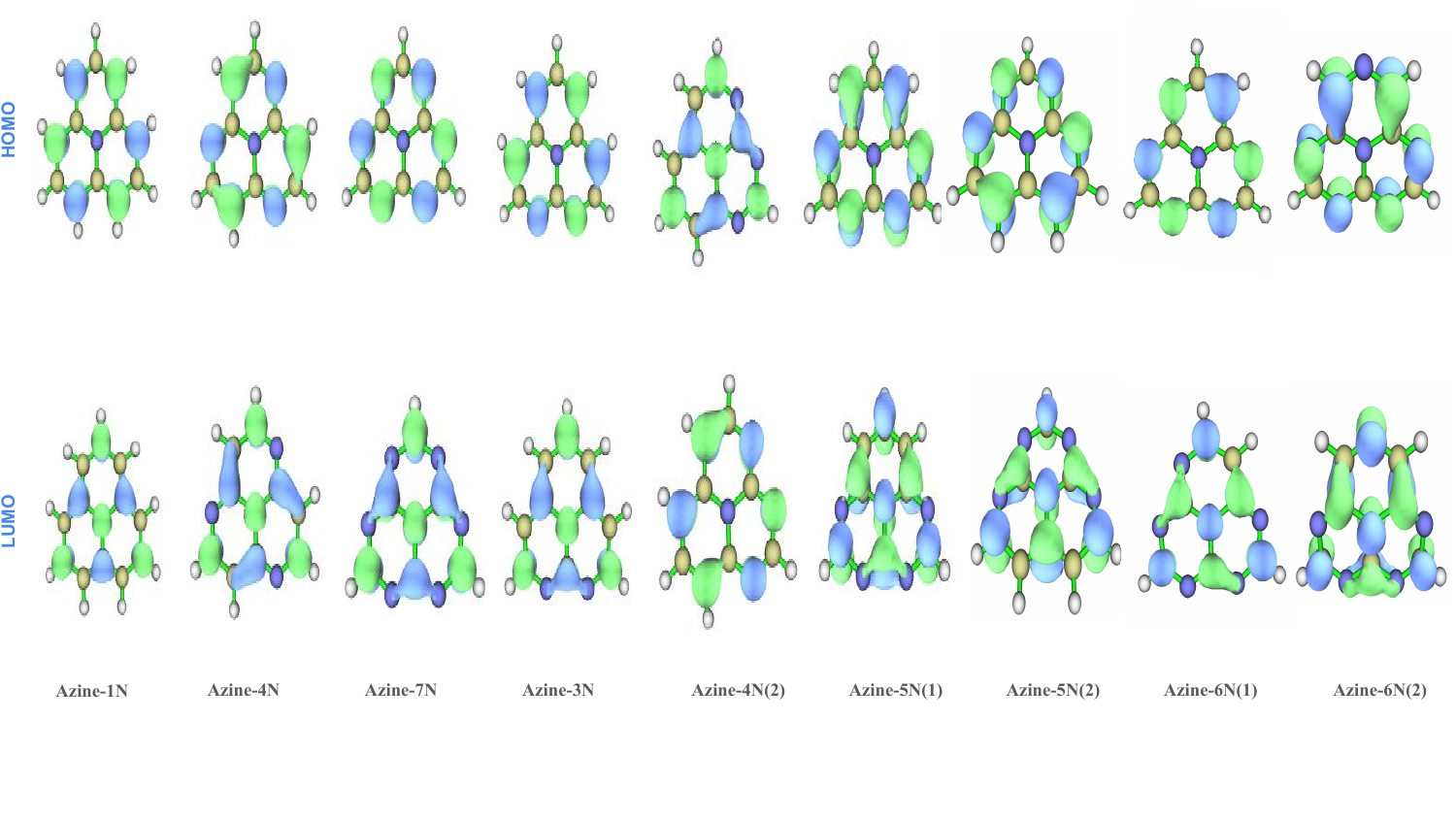}
  \caption{HOMO-LUMO of 9 molecules in Fig.~1 with B3LYP/def2-TZVP level of theory}
  \label{HOMO-LUMO}

\end{figure}

The singlet state's stabilisation has previously been linked to doubly excited singlet configurations that are energetically low-lying 
\cite{desilva2019b,ghosh2022,ricci2021}, which would indicate that capturing more electron correlation would lead us to a greater inversion. Certain forms of doubly excited determinants, 
are linked to what is called dynamic spin polarisation or the differential correlation of two states with same dominant orbital occupancy but different spin coupling. 
Recently it has been indirectly recognised as a critical effect in the ST gap inversion of the heptazine molecule\cite{drwal2023}. 
Although the quantitative estimation of spin polarisation is not feasible, but it is known to increase linearly as the highest occupied molecular orbital–lowest unoccupied molecular orbital (HOMO–LUMO) exchange integral (K$_{if}$) decreases and this integral is directly connected to the amplitude of the S$_1$–T$_1$ energy gap. The triangle-shaped topology and the significance of the sites of hetero-substitution with N/B atoms appears to influence the inversion and several studies have looked at this aspect\cite{ricci2021,dreuw2023}. In our test set, we have found that isomers with same chemical formula but different positions of N-atoms can show opposite signs of $\Delta$E$_{ST}$. However, by utilising the well-known Pariser–Parr–Pople model, which is a semi-empirical model for correlated electrons in $\pi$-conjugated systems, it has been reported by Bedogni et. al.\cite{bedogni2024}, that neither the triangle-shaped molecules nor any particular molecular symmetry are necessary for the ST inversion. Indeed, a small gap and a tiny exchange integral (K$_{if}$) between the frontier orbitals plays the deciding role. Further research is needed on which other structures can show this INVEST property and  large scale screening studies are underway even including machine-learning protocols\cite{barneschi2024,jorner2024}. Strictly non-overlapping HOMO and LUMO orbitals are not also a pre-requisite. When the electronic charge is transferred from the HOMO to the LUMO during the lowest energy transition, which has a dominant multi-resonant charge-transfer (MR-CT) character, the normally insignificant spin polarisation correction may be substantial enough to flip the ST gap as the exchange integral, which causes the ST splitting, is very small. We find, in this paper, that XC functionals designed to capture spin polarization do particularly well for these systems in the TDDFT framework, lending credence to the claims of \cite{drwal2023}.

However, apart from the traditional TDDFT, an alternative path to obtain open-shell singlet excited states at the SCF level is to use a modified algorithm to converge to a higher excited states. There are several strategies for this with the most common ones being the maximum overlap method (MOM)\cite{gilbert2008}, square-gradient minimization method (SGM)\cite{hait2020,hait2021}, restricted open-shell Kohn-Sham (ROKS)\cite{hait2016}, excited state DFT (e-DFT)\cite{edft2008} etc. These methods are said to express correct $1/R$ dependency at large R with standard functionals and accurately describe CT states at computational costs equivalent to typical ground-state DFT computations. Another approach that has been documented in the literature\cite{mei2019} uses the ground-state KS computation of the orbital energies of a $(N-1)$ electron system to calculate the excitation energy of an N electron system. However, this approach is only feasible if the ground and first singlet excited states have distinct symmetry. However, the $\Delta$SCF methods (MOM and SGM) suffer from severe spin contamination mostly due to forceful representation of a multi-determinant wave-function as a single-determinant. Moreover, non-orthogonality of ground and excited state functions makes it very difficult to obtain transition properties. Thus, we believe that $\Delta$SCF methods while cheap and convenient should be restricted to initial screening applications. MP2 and CCSD can also be applied on these optimized excited states leading to $\Delta$MP2 and $\Delta$CCSD excitation energies, which we already reported in\cite{chanda2023nature}.
In this paper, mainly MOM, SGM, and ROKS method are considered in the DFT framework. MOM \cite{gilbert2008} works by modifying the orbital selection step in the SCF procedure. Instead of filling orbitals in ascending order of their energies (aufbau protocol), MOM attempts to select occupied orbitals after each iteration by maximization of overlap with the occupied orbitals from the previous iteration thereby honing in on an electronic configuration of choice. For the ${S_1}$ state, we have adjusted the SCF process at each density matrix update step by maintaining the occupation numbers of $\beta$-HOMO and $\beta$-LUMO to 0 and 1, and $\alpha$-HOMO and $\alpha$-LUMO to 1 and 0, respectively. Similarly, for the $T_1$ state, we have fixed the occupation numbers of $\beta$-HOMO and $\beta$-LUMO to 0 and 0, and $\alpha$-HOMO and $\alpha$-LUMO 1 and 1, respectively. SGM\cite{hait2020} is an orbital optimization method which minimizes the square of the gradient of the Lagrangian $\nabla_\theta$$\mathcal{L}$
which is denoted as $\nabla_\theta$$\mathcal{L}$ and the square of gradient $\Delta$,
\begin{equation}
    \Delta=|\nabla_\theta\mathcal{L}|^2= \sum_{ai}\left|\frac{\partial\mathcal{L}}{\partial\theta_{ai}}\right|^2
\end{equation}

When used via a finite difference technique, the square gradient minimization (SGM) approach only needs analytical energy/Lagrangian orbital gradients and only costs three times as much as the ground state orbital optimisation (per iteration). To investigate the accuracy of orbital optimised DFT excited states, SGM is used to obtain both spin-purified restricted open-shell Kohn-Sham (ROKS) and spin broken single determinant S$_1$ states. It is discovered that SGM can converge difficult situations in which analogues or the maximal overlap method (MOM) either fail to converge or collapse to the ground state, but needs more intervention and trial and error.

In this framework, the inclusion of the dynamic spin-polarization is not quite straightforward. It is reflected in the extent of spin-contamination generated by various XC-functionals. We discuss these aspects in detail in Sec.~IVC.

\section{COMPUTATIONAL DETAILS}
We have chosen a group of N-substituted hydrocarbons that resemble triangles (Fig.~1) such as azine-1N, azine-4N and azine-7N. All molecular geometries are optimized in ORCA using B3LYP/def2-TZVP level of theory including Becke-Johnson dispersion corrections and are reported in the supplementary material. A trial version of Q-Chem 6.0.1 program package\cite{sato2015} has been used to carry out all the LR-TDDFT and $\Delta$SCF computations involving MOM, SGM, ROKS, $\Delta$ROSGM with the Local density approximation functional (LDA); BLYP, PBE and PB86 functionals at the generalized gradient approximation (GGA); TPSS and M06-L at meta-GGA level; B3LYP, PBE0 and BHHLYP at global hybrid GGA (GH-GGA); M06-2X, M08-HX, TPSSH at GH-meta-GGA level; $\omega$B97X, $\omega$B97X-V, $\omega$B97X-D3, CAMB3LYP, LRC-$\omega$PBEh at range separated hybrid GGA (RSH-GGA). TDDFT(D) calculations have been carried out in ORCA 5.0.4 software with double hybrid functionals namely PBE02, B2PLYP, PBE-QIDH, B2GP-PLYP, $\omega$PBEPP86, $\omega$B88PP86, $\omega$B2PLYP, SCS-PBEQIDH, SCS-$\omega$B88PP86 and SOS-$\omega$PBEPP86. All computations followed a strict SCF convergence limit of 10$^{-8}$. However, on encountering convergence difficulties with the MOM strategy, we decreased the convergence criteria to 10$^{-6}$.

\begin{table}[!htp]
\centering
\caption{Vertical excitation energies and associated $\Delta$E$_{ST}$ energy difference (all in eV) calculated with TDDFT. Geometries are optimized at B3LYP/def2-TZVP level}\label{tab:tddft}
\scriptsize
\begin{tabular}{|c |c c c|c c c|c c c|}
\toprule
Molecules && azine-1N &&& azine-4N &&& azine-7N & \\
Method &$S_1$ &$T_1$ &$\Delta$E$_{ST}$ &$S_1$ &$T_1$ &$\Delta$E$_{ST}$ & $S_1$ & $T_1$ &$\Delta$E$_{ST}$ \\
\hline
LDA &1.204 &1.037 &0.167 &2.086 &1.845 &0.241 &2.664 &2.485 &0.179 \\
\hline
\multicolumn{10}{|l|}{Generalized Gradient Approximation (GGA)} \\
\hline
BLYP &1.209 &1.033 &0.176 &2.090 &1.830 &0.260 &2.668 &2.479 &0.189 \\
BP86 &1.225 &1.037 &0.188 &2.113 &1.839 &0.274 &2.702 &2.503 &0.199 \\
PBE &1.223 &1.035 &0.188 &2.111 &1.837 &0.274 &2.698 &2.498 &0.200 \\
\hline
\multicolumn{10}{|l|}{Meta-Generalized Gradient Approximation (meta-GGA)} \\
\hline
M06-L &1.367 &1.183 &0.183 &2.308 &2.020 &0.289 &2.929 &2.726 &0.203 \\
TPSS &1.271 &1.078 &0.193 &2.187 &1.898 &0.289 &2.789 &2.584 &0.205 \\
\hline
\multicolumn{10}{|l|}{Global Hybrid Generalized Gradient Approximation (GH GGA)} \\
\hline
B3LYP &1.262 &1.065 &0.197 &2.283 &1.963 &0.320 &2.942 &2.723 &0.219 \\
PBE0 &1.292 &1.072 &0.220 &2.355 &1.996 &0.359 &3.042 &2.799 &0.244 \\
BHHLYP &1.395 &1.160 &0.234 &2.618 &2.181 &0.438 &3.409 &3.136 &0.273 \\
\hline
\multicolumn{10}{|l|}{Global Hybrid Meta-Generalized Gradient Approximation (GH meta-GGA)} \\
\hline
TPSSH &1.292 &1.088 &0.204 &2.277 &1.956 &0.321 &2.918 &2.697 &0.221 \\
M08-HX &1.343 &1.175 &0.167 &2.500 &2.184 &0.316 &3.158 &2.964 &0.194 \\
M06-2X &1.312 &1.125 &0.188 &2.502 &2.170 &0.332 &3.186 &2.969 &0.216 \\
\hline
\multicolumn{10}{|l|}{Range-Separated Hybrid Generalized Gradient Approximation (RSH GGA)} \\
\hline
$\omega$B97x &1.383 &1.163 &0.221 &2.625 &2.201 &0.424 &3.330 &3.078 &0.252 \\
$\omega$B97X-V &1.393 &1.177 &0.216 &2.636 &2.227 &0.409 &3.346 &3.099 &0.247 \\
$\omega$B97X-D3 &1.366 &1.151 &0.215 &2.575 &2.175 &0.400 &3.278 &3.033 &0.245 \\
CAMB3LYP &1.320 &1.099 &0.221 &2.485 &2.088 &0.397 &3.173 &2.924 &0.249 \\
LRC-$\omega$PBEh &1.330 &1.092 &0.238 &2.492 &2.078 &0.415 &3.180 &2.916 &0.264 \\
\hline
\multicolumn{10}{|l|}{Double Hybrid Generalized Gradient Approximation (DH GGA)} \\
\hline
$\omega$B97X-2 & 0.719 & 1.353 & -0.634 &  1.721 & 2.689 & -0.968 &  1.860 & 2.631 & -0.771 \\
PBE0-2 & 1.113 & 1.211 & -0.098 &  2.323 & 2.330 & -0.007  & 2.884 & 3.047 & -0.163 \\
B2PLYP & 1.116 & 1.099 & 0.017  & 2.191 & 2.015 & 0.176 &  2.755 & 2.790 & -0.035 \\
$\omega$B2PLYP & 1.211 & 1.091 & 0.120  & 2.473 & 2.069 & 0.404  & 3.112 & 3.046 & 0.066 \\
SCS/SOS-$\omega$B2PLYP & 1.047 & 1.064 & -0.017  & 2.327 & 2.063 & 0.264 & 2.887 & 3.002 & -0.115 \\
B2GP-PLYP & 1.117 & 1.142 & -0.025  & 2.252 & 2.121 & 0.131 &  2.817 & 2.906 & -0.089 \\
$\omega$B2GP-PLYP & 1.188 & 1.136 & 0.052 &  2.452 & 2.144 & 0.308  & 3.074 & 3.083 & -0.009 \\
SCS-$\omega$B2GP-PLYP & 0.936 & 0.996 & -0.060 &  2.215 & 2.031 & 0.184 &  2.742 & 2.931 & -0.189 \\
SOS-$\omega$B2GP-PLYP & 0.961 & 1.079 & -0.118  & 2.249 & 2.118 & 0.131  & 2.771 & 3.013 & -0.242 \\
PBE-QIDH & 1.176 & 1.135 & 0.041 &  2.369 & 2.108 & 0.261 &  2.994 & 3.008 & -0.014 \\
SCS-PBEQIDH & 1.028 & 1.151 & -0.123  & 2.225 & 2.160 & 0.065  & 2.757 & 2.974 & -0.217 \\
SOS-PBEQIDH & 0.999 & 1.159 & -0.16 & 2.203 & 2.173 & 0.03  & 2.719 & 2.985 & -0.266 \\
$\omega$PBEPP86 & 1.110 & 1.172 & -0.062 &  2.310 & 2.242 & 0.068  & 2.863 & 2.995 & -0.132 \\
SCS-$\omega$PBEPP86 & 0.869 & 0.873 & -0.004 & 2.109 & 1.908 & 0.201 & 2.637 & 2.789 & -0.152 \\
SOS-$\omega$PBEPP86 & 0.885 & 1.015 & -0.130 &  2.143 & 2.065 & 0.078 &  2.650 & 2.927 & -0.277 \\
$\omega$B88PP86 & 1.135 & 1.142 & -0.007 &  2.338 & 2.168 & 0.170 &  2.912 & 2.985 & -0.073 \\
SCS-$\omega$B88PP86 & 1.016 & 1.081 & -0.065  & 2.261 & 2.096 & 0.165  & 2.804 & 2.975 & -0.171 \\
SOS-$\omega$B88PP86 & 0.964 & 1.059 & -0.095 &  2.213 & 2.079 & 0.134  & 2.739 & 2.959 & -0.220 \\ 
\hline
NEVPT2(12,9) & 0.964 & 1.055 & -0.091 &  2.246 & 2.231 & 0.015 &  2.59 & 2.77 & -0.18  \\
EOM-CCSD & 1.068 & 1.146 & -0.077 &  2.382 & 2.213 & 0.169  & 2.916 & 3.066 & -0.150  \\
FICMRCI(12,9)\cite{chanda2023nature} &1.022 &1.099 &-0.077 &2.249 &2.235 &0.014 &2.83 &2.995 &-0.165 \\
\hline
\end{tabular}
\end{table}

\section{Results and Discussion}
In our first, most extensive study, we have looked at the electronic structures of the three template molecules - azine-1N, 4N and 7N in their S$_0$, $S_1$, and $T_1$ states. The experiment determined the singlet–triplet gaps for azine-1N and azine-4N to be 0.08 and $<$0.1 eV, respectively \cite{jacs80,jacs86}. Azine-7N or heptazine, has received a lot of attention in recent years because of its theoretically predicted large inverted singlet-triplet gap of roughly -0.25 eV\cite{desilva2019a}. We shortlist the method and functional pairs with the best performance for S$_1-$S$_0$, T$_1-$S$_0$, and $\Delta$E$_{ST}$ and apply them next to 6 other related templates shown in Fig.~1 to further evaluate them. While the test set is not huge, our focus has been to cover a large number of functionals and understand the theoretical underpinings for their performance rather than base our recommendations on statistical errors only. At the end of the study, we make our recommendations along with the theoretical justifications. Attempts have been made to explain the performance of the functionals at various rungs of Jacob's ladder. The existence of inverted singlet–triplet gaps in these molecules was theoretically confirmed by our EOM-CCSD and FICMRCISD results recently\cite{chanda2023nature}. The ST gap with EOM-CCSD for azine-4N is 0.169 eV, a bit away from experiment ($<$0.1 eV) but FICMRCISD is on point (0.014 eV with (12,9) CAS and 0.007 eV for (14,10) CAS). Fig.~2 shows the HOMO and LUMO of these molecules, with clearly visible spatially separated electron densities. This reveals the charge-transfer (CT) character with nearly degenerate excited singlet and triplet states\cite{desilva2019a}. Minimizing the singlet-triplet gap through long-ranged HOMO-LUMO separations is merely one of the determining factors in the design of these molecules. The present study showcases the results of inverted singlet-triplet gaps tending to vanish and approach zero in the $\Delta$SCF-DFT framework utilizing MOM or SGM and TDDFT(D) with proper choice of functionals. Some of the molecules in the test set also show positive ST gaps to serve as counter examples for avoiding bias. Although S$_1-$S$_0$ and T$_1-$S$_0$ dictates the $\Delta$E$_{ST}$ value, the errors in the latter has been additionally included in our average error estimates to give additional weightage to methods with good error cancellation.

\begin{sidewaystable} 
\caption{S$_1$, T$_1$ and $\Delta$E$_{ST}$ (all in eV) calculated with TDDFT for 6 other templates. Geometries are optimized with B3LYP/def2-TZVP}\label{tab: }
\centering
\scriptsize
\begin{tabular}{|c|cccccccccccccccccc|}
\hline	
Molecules & & azine-3N & & & azine-4N(2) & & & azine-5N(1) & & & azine-5N(2) & & & azine-6N(1) & & & azine-6N(2) & \\
\hline
Method & $S_1$ & $T_1$ & $\Delta E_{ST}$ & $S_1$ & $T_1$ & $\Delta E_{ST}$ & $S_1$ & $T_1$ & $\Delta E_{ST}$ & $S_1$ & $T_1$ & $\Delta E_{ST}$ & $S_1$ & $T_1$ & $\Delta E_{ST}$ & $S_1$ & $T_1$ & $\Delta E_{ST}$ \\
\hline
LDA & 1.6896 & 1.4784 & 0.2112 & 1.970 & 1.735 & 0.235 & 2.186 & 1.966 & 0.22 & 2.164 & 1.935 & 0.229 & 2.412 & 2.207 & 0.205 & 2.113 & 1.812 & 0.301 \\
BLYP & 1.6948 & 1.4713 & 0.2235 & 1.974 & 1.724 & 0.250 & 2.191 & 1.957 & 0.234 & 2.168 & 1.925 & 0.243 & 2.416 & 2.198 & 0.218 & 2.113 & 1.791 & 0.322 \\
BP86 & 1.7172 & 1.4809 & 0.2363 & 1.999 & 1.7364 & 0.263 & 2.219 & 1.972 & 0.247 & 2.196 & 1.941 & 0.255 & 2.447 & 2.217 & 0.23 & 2.140 & 1.803 & 0.337 \\
PBE & 1.7146 & 1.4781 & 0.2365 & 1.996 & 1.733 & 0.263 & 2.216 & 1.969 & 0.247 & 2.193 & 1.937 & 0.256 & 2.444 & 2.2138 & 0.2302 & 2.138 & 1.801 & 0.337 \\
M06-L & 1.8871 & 1.6454 & 0.2417 & 2.182 & 1.908 & 0.274 & 2.411 & 2.156 & 0.255 & 2.393 & 2.127 & 0.266 & 2.657 & 2.419 & 0.238 & 2.322 & 1.955 & 0.367 \\
TPSS & 1.8871 & 1.6454 & 0.2417 & 2.067 & 1.792 & 0.275 & 2.294 & 2.037 & 0.257 & 2.270 & 2.002 & 0.268 & 2.528 & 2.289 & 0.239 & 2.206 & 1.845 & 0.361 \\
B3LYP & 1.8179 & 1.5572 & 0.2607 & 2.138 & 1.84 & 0.298 & 2.389 & 2.108 & 0.281 & 2.362 & 2.068 & 0.294 & 2.651 & 2.39 & 0.261 & 2.283 & 1.88 & 0.403 \\
PBE0 & 1.8717 & 1.5819 & 0.2898 & 2.205 & 1.872 & 0.333 & 2.466 & 2.152 & 0.314 & 2.438 & 2.11 & 0.328 & 2.74 & 2.447 & 0.293 & 2.353 & 1.903 & 0.450 \\
BHHLYP & 2.0568 & 1.7269 & 0.3299 & 2.435 & 2.041 & 0.394 & 2.736 & 2.357 & 0.379 & 2.704 & 2.311 & 0.393 & 3.056 & 2.706 & 0.35 & 2.580 & 2.008 & 0.572 \\
TPSSH & 1.8329 & 1.5674 & 0.2655 & 2.143 & 1.841 & 0.302 & 2.386 & 2.104 & 0.282 & 2.360 & 2.065 & 0.295 & 2.638 & 2.376 & 0.262 & 2.285 & 1.882 & 0.403 \\
M08-HX & 1.9395 & 1.7069 & 0.2326 & 2.294 & 2.018 & 0.276 & 2.559 & 2.302 & 0.257 & 2.526 & 2.252 & 0.274 & 2.847 & 2.605 & 0.242 & 2.449 & 2.069 & 0.380 \\
M06-2X & 1.9306 & 1.6705 & 0.2601 & 2.295 & 1.991 & 0.304 & 2.571 & 2.284 & 0.287 & 2.535 & 2.229 & 0.306 & 2.865 & 2.596 & 0.269 & 2.447 & 2.028 & 0.419 \\
$\omega$B97x & 2.0321 & 1.7237 & 0.3084 & 2.404 & 2.035 & 0.369 & 2.692 & 2.338 & 0.354 & 2.658 & 2.291 & 0.367 & 3.001 & 2.674 & 0.327 & 2.527 & 1.988 & 0.539 \\
$\omega$B97X-V & 2.0443 & 1.7412 & 0.3031 & 2.417 & 2.055 & 0.362 & 2.706 & 2.36 & 0.346 & 2.673 & 2.312 & 0.361 & 3.017 & 2.697 & 0.32 & 2.541 & 2.013 & 0.528 \\
$\omega$B97X-D3 & 2.0029 & 1.704 & 0.2989 & 2.369 & 2.014 & 0.355 & 2.652 & 2.313 & 0.339 & 2.619 & 2.266 & 0.353 & 2.954 & 2.641 & 0.313 & 2.496 & 1.985 & 0.511 \\
CAMB3LYP & 1.9353 & 1.6362 & 0.2991 & 2.291 & 1.939 & 0.352 & 2.565 & 2.231 & 0.334 & 2.186 & 2.185 & 0.001 & 2.857 & 2.547 & 0.31 & 2.424 & 1.931 & 0.493 \\
LRC-$\omega$PBEh & 1.9466 & 1.6299 & 0.3167 & 2.301 & 1.932 & 0.369 & 2.576 & 2.226 & 0.35 & 2.543 & 2.178 & 0.365 & 2.867 & 2.54 & 0.327 & 2.436 & 1.927 & 0.509 \\
\hline
$\omega$B97X-2 & 1.129 & 1.808 & -0.679 & 1.427 & 2.328 & -0.901 & 1.591 & 2.454 & -0.863 & 1.516 & 2.304 & -0.788 & 1.727 & 2.639 & -0.912 & 1.537 & 2.683 & -1.146 \\
PBE0-2 & 1.706 & 1.754 & -0.048 & 2.070 & 2.21 & -0.140 & 2.331 & 2.431 & -0.1 & 2.281 & 2.327 & -0.046 & 2.6 & 2.756 & -0.156 & 2.188 & 2.244 & -0.056 \\
B2PLYP & 1.657 & 1.6 & 0.057 & 1.987 & 1.912 & 0.075 & 2.232 & 2.197 & 0.035 & 2.190 & 2.133 & 0.057 & 2.484 & 2.474 & 0.01 & 2.108 & 1.993 & 0.115 \\
$\omega$B2PLYP & 1.842 & 1.652 & 0.19 & 2.216 & 1.976 & 0.240 & 2.493 & 2.301 & 0.192 & 2.455 & 2.231 & 0.224 & 2.796 & 2.636 & 0.16 & 2.316 & 1.939 & 0.377 \\
SCS/SOS-$\omega$B2PLYP & 1.688 & 1.618 & 0.07 & 2.038 & 1.941 & 0.097 & 2.301 & 2.26 & 0.041 & 2.261 & 2.187 & 0.074 & 2.592 & 2.595 & -0.003 & 2.12 & 1.876 & 0.244 \\
B2GP-PLYP & 1.681 & 1.664 & 0.017 & 2.026 & 2.009 & 0.017 & 2.278 & 2.296 & -0.018 & 2.233 & 2.217 & 0.016 & 2.539 & 2.589 & -0.05 & 2.145 & 2.085 & 0.060 \\
$\omega$B2GP-PLYP & 1.816 & 1.697 & 0.119 & 2.191 & 2.045 & 0.146 & 2.467 & 2.359 & 0.108 & 2.425 & 2.278 & 0.147 & 2.764 & 2.692 & 0.072 & 2.293 & 2.03 & 0.263 \\
SCS-$\omega$B2GP-PLYP & 1.549 & 1.552 & -0.003 & 1.919 & 1.899 & 0.02 & 2.176 & 2.207 & -0.031 & 2.133 & 2.121 & 0.012 & 2.458 & 2.543 & -0.085 & 1.999 & 1.821 & 0.178 \\
SOS-$\omega$B2GP-PLYP & 1.576 & 1.632 & -0.056 & 1.947 & 1.981 & -0.034 & 2.203 & 2.285 & -0.082 & 2.161 & 2.2 & -0.039 & 2.487 & 2.621 & -0.134 & 2.024 & 1.917 & 0.107 \\
PBE-QIDH & 1.778 & 1.678 & 0.1 & 2.139 & 2.027 & 0.112 & 2.409 & 2.33 & 0.079 & 2.366 & 2.249 & 0.117 & 2.693 & 2.647 & 0.046 & 2.259 & 2.057 & 0.202 \\
SCS-PBEQIDH & 1.609 & 1.678 & -0.069 & 1.964 & 2.034 & -0.070 & 2.216 & 2.318 & -0.102 & 2.170 & 2.232 & -0.062 & 2.482 & 2.631 & -0.149 & 2.068 & 2.048 & 0.020 \\
SOS-PBEQIDH & 1.579 & 1.685 & -0.106 & 1.934 & 2.04 & -0.106 & 2.183 & 2.323 & -0.14 & 2.136 & 2.236 & -0.1 & 2.447 & 2.638 & -0.191 & 2.032 & 2.045 & -0.013 \\
$\omega$PBEPP86 & 1.698 & 1.712 & -0.014 & 2.058 & 2.107 & -0.049 & 2.316 & 2.377 & -0.061 & 2.267 & 2.281 & -0.014 & 2.583 & 2.685 & -0.102 & 2.171 & 2.168 & 0.003 \\
SCS-$\omega$PBEPP86 & 1.468 & 1.429 & 0.039 & 1.832 & 1.787 & 0.045 & 2.087 & 2.091 & -0.004 & 2.042 & 1.999 & 0.043 & 2.359 & 2.423 & -0.064 & 1.925 & 1.708 & 0.217 \\
SOS-$\omega$PBEPP86 & 1.486 & 1.566 & -0.08 & 1.851 & 1.93 & -0.079 & 2.103 & 2.223 & -0.12 & 2.057 & 2.132 & -0.075 & 2.375 & 2.557 & -0.182 & 1.934 & 1.87 & 0.064 \\
$\omega$B88PP86 & 1.729 & 1.683 & 0.046 & 2.089 & 2.053 & 0.036 & 2.351 & 2.338 & 0.013 & 2.305 & 2.25 & 0.055 & 2.625 & 2.65 & -0.025 & 2.200 & 2.091 & 0.109 \\
SCS-$\omega$B88PP86 & 1.62 & 1.624 & -0.004 & 1.984 & 1.973 & 0.011 & 2.351 & 2.338 & 0.013 & 2.198 & 2.185 & 0.013 & 2.52 & 2.599 & -0.079 & 2.076 & 1.932 & 0.144 \\
SOS-$\omega$B88PP86 & 1.565 & 1.604 & -0.039 & 1.929 & 1.955 & -0.026 & 2.184 & 2.252 & -0.068 & 2.14 & 2.166 & -0.026 & 2.46 & 2.582 & -0.122 & 2.015 & 1.907 & 0.108 \\
\hline
NEVPT2(12,9) & 1.495 & 1.611 & -0.116 & 2.013 & 2.012 & 0.001 & 2.543 & 2.511 & 0.032 & 2.137 & 2.178 & -0.041 & 2.508 & 2.605 & -0.097 & 2.374 & 2.016 & 0.358 \\
EOM-CCSD & 1.7033 & 1.7274 & -0.0241 & 2.054 & 2.0421 & 0.0119 & 2.342 & 2.3714 & -0.0298 & 2.2454 & 2.2712 & -0.0258 & 2.5717 & 2.6349 & -0.0632 & 2.155 & 2.0765 & 0.0785 \\
\hline
\end{tabular}
\end{sidewaystable}

\subsection{Evaluation of TDDFT and TDDFT(D) performance}
TDDFT generally offers data with a fair amount of precision at affordable computing costs, according to a recent thorough examination of the accuracy of TDDFT calculations using 43 functionals on a significant benchmark set \cite{Liang2022}. Owing to the disjoint nature of the HOMO and LUMO, as well as the fact that excited states in the single-excitation picture are primarily HOMO to LUMO excitations, these excitations could thus be viewed as intramolecular (short-range) charge-transfer excitations. Previous failures of TD-DFT for dealing with charge-transfer excitations were corrected using a long-range correction to the XC functional, of which the $\omega$B97XD model\cite{chai2008} was one of the most successfully used versions. Nevertheless, the application of this functional to e.g. azine-7N in previous publications\cite{ehrmaier2019} also produced a positive ST gap of 0.23 ev, indicating that the incorrect prediction of the negative ST values is not due to short- vs. long-range effects of the exchange-correlation potential. The singlet-triplet inversion of the INVEST molecules, in general, could not be accurately predicted by TDDFT primarily due to poor S$_1$ state energies. The T$_1-$S$_0$  energies are quite stable across various functionals (with a few outliers of course). The data in Table 1 demonstrates that, unless double hybrid functionals (DH GGA) are used, TDDFT anticipates a positive singlet-triplet energy gap in the range of [0.167, 0.424] for azine-1N, 4N and 7N. The $\Delta$E$_{ST}$  values for pure functionals (LDA, GGA, meta GGA) are smaller, whereas the inclusion of exact exchange in hybrid functionals (GH GGA, GH meta-GGA) tends to increase the singlet-triplet energy separation. The hybrid functionals in Table~I are arranged roughly in increasing HFX with GH, RSH and spin-component scaled versions (SCS and/or SOS) grouped together. The increase in the $\Delta$E$_{ST}$ gap can be explained by the fact that high-spin states are generally stabilized relative to low-spin ones by the incorporation of exact exchange. Note that $\Delta$E$_{ST}$ values are lower for GH-meta GGA Minnesota functionals than for GH-GGA inspite of larger amount of HF exchange, because the former have more advanced empirical parameterization that takes into account more correlations in addition to exchange. In these molecules electron correlations in S$_1$ negate the stability from exchange interactions bringing S$_1$ and T$_1$ closer. However, being empirical, the trends are not very clean. One may see (Fig.~3) that except for the Minnesota functionals, positivity of $\Delta$E$_{ST}$ correlates well with the \%HF exchange. RSH GGA functionals have even more HFX leading to S$_1-$S$_0$ being further pushed up. 

\begin{figure}
    \centering
    \includegraphics[width=1.0\linewidth]{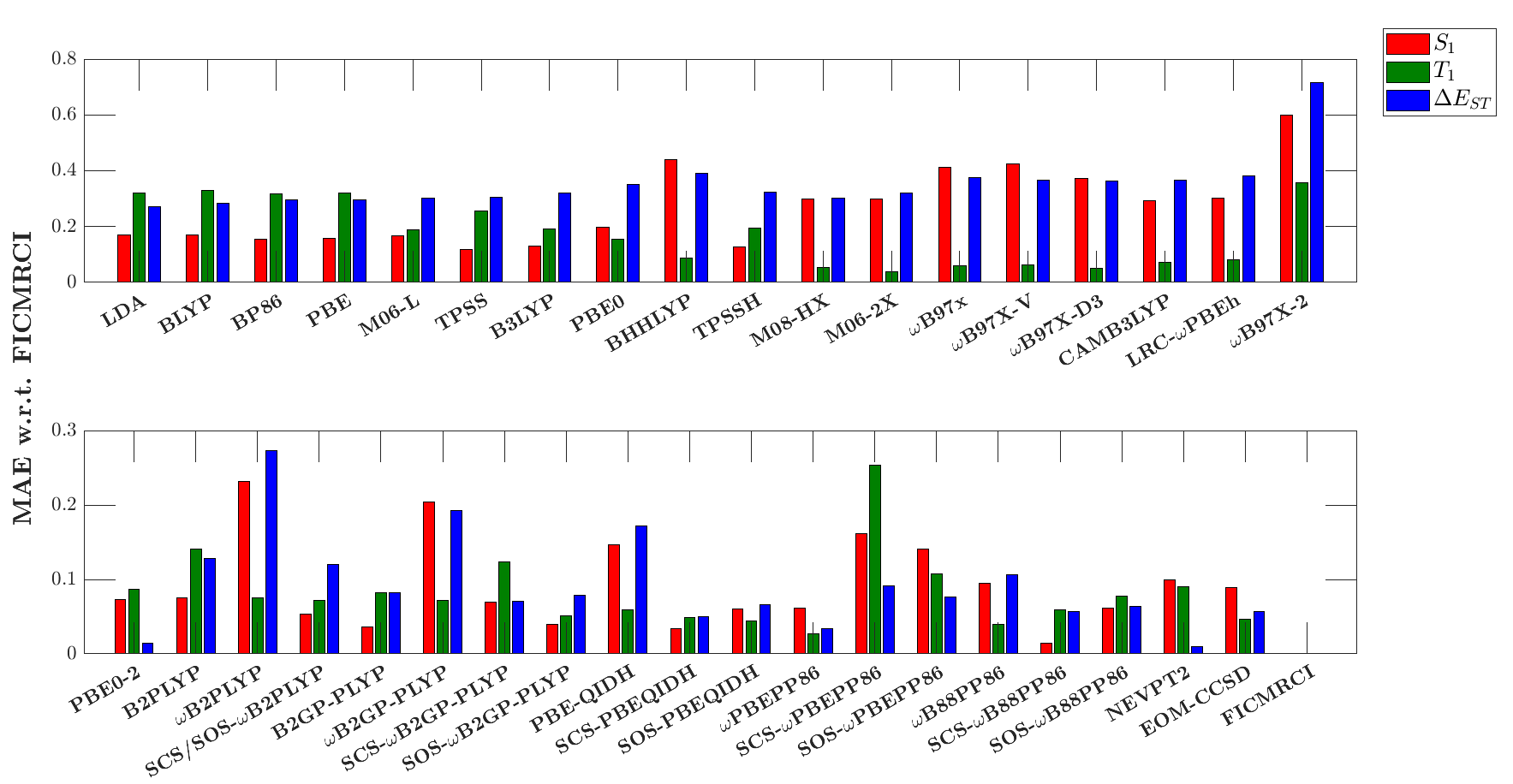}
\caption{Mean absolute error (MAE) of $S_{1}-S_{0}$, $T_{1}-S_{0}$,  and 
 $\Delta$E$_{ST}$ energy values in eV between TDDFT with different methods and the ICMRCI (12,9) for (a) azine-1N, (b) azine-4N and (c) azine-7N}
\label{tddft_fig}   
\end{figure}

In the case of DH GGA functionals using ORCA 5.0.4, PBE0-2, PBE-QIDH, and B2PLYP yielded mixed degree of success in predicting the sign of $\Delta$E$_{ST}$ for various molecules. $\omega$B2PLYP functional yields positive $\Delta$E$_{ST}$ values and $\omega$B97X-2 predicted negative $\Delta$E$_{ST}$ gaps for every molecule, although it consistently overestimated their $\Delta$E$_{ST}$ values. All other DH functionals studied by us give inverted or near degenerate $S_1$ and $T_1$ energies. $\omega$PBEPP86, SCS-$\omega$B88PP86, SOS-$\omega$PBEPP86, and SCS-PBE-QIDH functionals predicted an almost zero or negative $\Delta$E$_{ST}$ gap for all molecules. In particular, $\omega$PBEPP86 and SCS-$\omega$B88PP86 are the most accurate and robust methods for obtaining excitation energies, and $\Delta$E$_{ST}$ gap. A comparison between the unscaled and SCS versions of the DH functionals in Table~1 clearly indicates that spin-component scaling (SCS) consistently moves S$_1$, T$_1$ exciation energies and $\Delta$E$_{ST}$ towards the FIC-MRCISD values. The SCS versions consistently do better than the spin-opposite scaled (SOS) variants pointing towards dynamic spin-polarization being the dominant physical effect for these molecules and supports the findings of Drwal et al\cite{drwal2023} and Alipour et al\cite{alipour2022}. The only exception is $\omega$PBEPP86 which performs very well with a MAE of 0.04 but deteriorates to 0.11 with SOS and 0.17 with SCS techniques. We do not understand the reason for this.

We thus see that, the lack of state-specific correlation and spin polarization in LR-TDDFT can be corrected by a post-KS-TDDFT state-specific MP2-like correction to the excited state energies in TDDFT(D). The mean absolute error (MAE) between the LR-TDDFT and ICMRCI (12,9) results of the three molecules for S$_1$, T$_1$ and $\Delta$E$_{ST}$ are displayed in Fig.~3. 
The average absolute error in $S_1$-$S_0$, $T_1$-$S_0$ and $\Delta$E$_{ST}$ is less than 0.1eV for all three molecules in our primary set when PBE0-2, SCS-PBEQIDH, B2GP-PLYP, $\omega$PBEPP86, $\omega$B88PP86 and SCS-$\omega$B88PP86 functionals are used. Among these, SCS-PBEQIDH, $\omega$PBEPP86 and SCS-$\omega$B88PP86 have the least average error of 0.04eV. The S$_1$, T$_1$ and $\Delta$E$_{ST}$ values for secondary test set of molecules are given in Table~II. The total set of 9 molecules are benchmarked against EOM-CCSD values. The signed errors in the individual excitation energies for both primary and secondary test set of molecules are given in detail in the Table~S1 and S2 in the supplementary material.  
Based on the above analysis, the XC functionals with $MAE < 0.1 eV$ for the primary and secondary test sets are $\omega$PBEPP86 $<$ SCS-$\omega$B88PP86 $<$ SCS-PBEQIDH $<$ PBE0-2 $<$ B2GP-PLYP $<$ $\omega$B88PP86 ordered in ascending average error. SOS-$\omega$PBEPP86, B2PLYP and PBE-QIDH are also close in performance. $\omega$B2PLYPand $\omega$B97X-2 have the maximum MAE. 

\begin{figure}
\centering
    \includegraphics[width=1.0\linewidth]{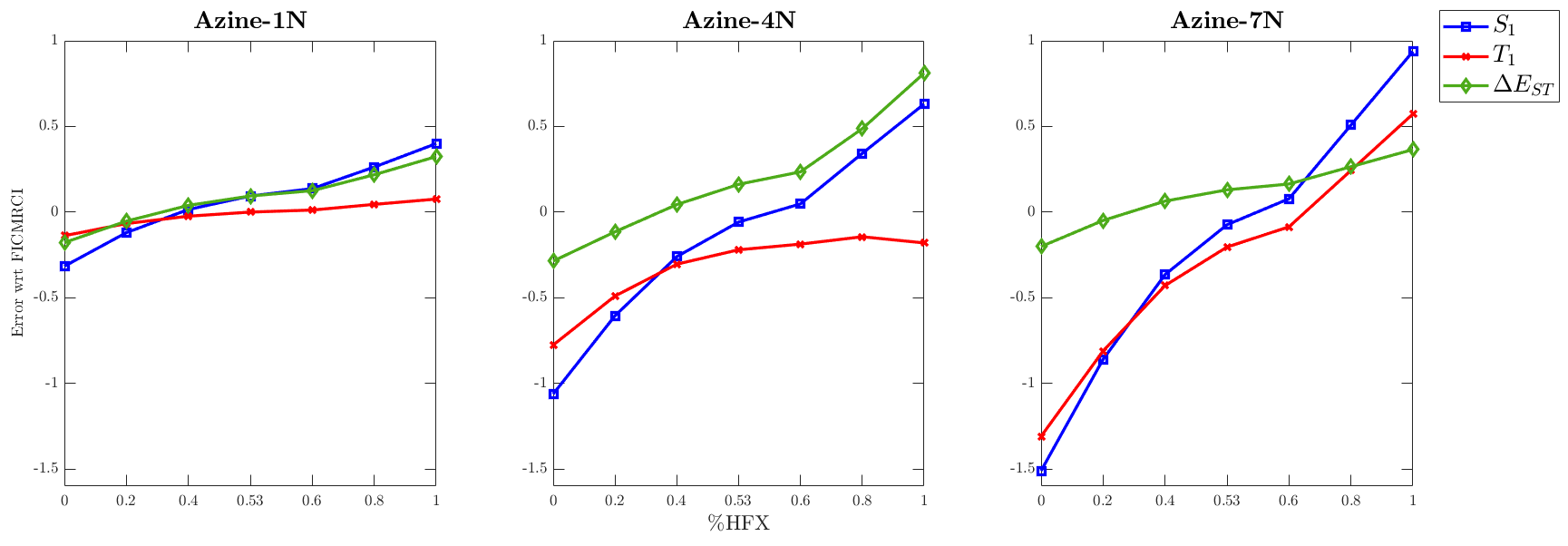}
\caption{Variation of errors in S$_1$, T$_1$ and $\Delta$E$_{ST}$ for TDDFT-B2PLYP/def2-TZVP w.r.t. FICMRCI with \%HF exchange for azine-1N, 4N and 7N.}
\label{hfex_tddft_fig}   
\end{figure}

\begin{table}[h]\centering
\caption{Singlet-Triplet gap as a fractional variation of HF exchange for TDDFT-B2PLYP}\label{tab: }
\scriptsize
\begin{tabular}{|c|c|ccc|ccc|ccc|}
\hline
 & & & azine-1N & & & azine-4N & & & azine-7N & \\
\hline
HF Exchange & B88 Exchange & S$_1$ & T$_1$ & $\Delta$E$_{ST}$ & S$_1$ & T$_1$ & $\Delta$E$_{ST}$ & S$_1$ & T$_1$ & $\Delta$E$_{ST}$ \\
\hline
0 & 1 & 0.706 & 0.961 & -0.255 & 1.187 & 1.459 & -0.272 & 1.319 & 1.685 & -0.366 \\
0.2 & 0.8 & 0.902 & 1.032 & -0.13 & 1.644 & 1.745 & -0.101 & 1.969 & 2.183 & -0.214 \\
0.4 & 0.6 & 1.037 & 1.075 & -0.038 & 1.989 & 1.931 & 0.058 & 2.464 & 2.565 & -0.101 \\
\textbf{0.53} & \textbf{0.47} & \textbf{1.116} & \textbf{1.099} & \textbf{0.017} & \textbf{2.191} & \textbf{2.015} & \textbf{0.176} & \textbf{2.755} & \textbf{2.79} & \textbf{-0.035} \\
0.6 & 0.4 & 1.159 & 1.111 & 0.048 & 2.297 & 2.048 & 0.249 & 2.907 & 2.908 & -0.001 \\
0.8 & 0.2 & 1.285 & 1.144 & 0.141 & 2.592 & 2.091 & 0.501 & 3.337 & 3.237 & 0.1 \\
1 & 0 & 1.422 & 1.175 & 0.247 & 2.88 & 2.055 & 0.825 & 3.767 & 3.567 & 0.2 \\
\hline
\end{tabular} 
\end{table}

It has been noted that the LR-TDDFT theory with standard XC kernels consistently yields positive STGs for azine-7N independent of the amount of non-local HFX\cite{bhattacharyya2021,ricci2021,tuckova2022}. 
To explore the relation of \%HFX and $\Delta$E$_{ST}$ directly, 
the variation of $\Delta$E$_{ST}$ for TDDFT with B2PLYP as a function of increasing \%HF exchange is presented in Table~III. The fraction of DFT exchange is also adjusted to make total exchange unity. It has been observed that for all cases ST gap increases with the amount of HFX starting from a negative value, it eventually goes to a positive ST gap. Fig.~4 represents the variation of errors in S$_1$, T$_1$ and $\Delta$E$_{ST}$ with respect to FICMRCI for the three molecules azine-1N, 4N and 7N. Error in S$_1$ changes abruptly for azine-4N and 7N while for T$_1$ it changes smoothly leading to a smooth change of error in ST gap with increasing \%HFX. With reference to the Table~S1 in the supplementary material, it may be noted that among the GH GGA functionals errors increase with the \%HFX. On the other hand, the empirically parameterized GH meta-GGAs (M06-2X and M08-HX) with more than 50\% HFX perform similar to TPSSH with HFX=10\%. TPSSH seems to do well for the S$_1-$S$_0$ gap but consistently underestimates the T$_1-$S$_0$ gap. RSH functionals with 100\% LR-HFX show larger errors. Since, the leading contribution to $\Delta$E$_{ST}$ is the HOMO-LUMO exchange integral K$_{if}$, it stands to reason that the amount of HFX in the XC functionals correlates with the prediction accuracy of $\Delta$E$_{ST}$ in INVEST molecules especially because the exchange and correlation interactions are delicately balanced in these systems and stabilization of T$_1$ by exchange needs a large amount of correlation to offset it. Bruckner and Engels\cite{bruckner2017} have made an exhaustive study of similar correlations in diverse classes of molecules aimed at screening candidates for singlet fission, but one may extend their arguments to INVEST molecules as well.

\begin{table}[!htp]\centering
\caption{Vertical excitation energies and associated $\Delta$E$_{ST}$ energy difference (all in eV) calculated with MOM and SGM. Geometries are optimized at B3LYP/def2-TZVP level}\label{tab: }
\scriptsize
\begin{tabular}{|l|cccc|cccc|cccc|}\toprule
Molecules &&& azine-1N &&&&azine-4N &&&&azine-7N&  \\
\hline
Method &S1 &T1 &\parbox{1.5cm}{$\Delta$E$_{ST}$\\(MOM)} &\parbox{1.5cm}{$\Delta$E$_{ST}$\\(SGM)} &S1 &T1 &\parbox{1.5cm}{$\Delta$E$_{ST}$\\(MOM)} &\parbox{1.5cm}{$\Delta$E$_{ST}$\\(SGM)} &S1 &T1 &\parbox{1.5cm}{$\Delta$E$_{ST}$\\(MOM)} &\parbox{1.5cm}{$\Delta$E$_{ST}$\\(SGM)} \\
\hline
LDA &1.149 &1.125 &0.024 &0.024 &1.977 &1.930 &0.047 &0.047 &2.598 &2.582 &0.016 &0.016 \\
\hline
\multicolumn{13}{|l|}{Generalized Gradient Approximation (GGA)} \\
\hline
BLYP &1.061 &1.073 &-0.012 &-0.012 &1.901 &1.878 &0.023 &0.023 &2.507 &2.522 &-0.023 &-0.015 \\
BP86 &1.061 &1.074 &-0.012 &-0.013 &1.912 &1.887 &0.025 &0.025 &2.521 &2.546 &-0.024 &-0.025 \\
PBE &1.063 &1.073 &-0.010 &-0.009 &1.913 &1.886 &0.028 &0.028 &2.522 &2.543 &-0.021 &-0.021 \\
\hline
\multicolumn{13}{|l|}{Meta-Generalized Gradient Approximation (meta-GGA)} \\
\hline
M06-L &1.045 &1.096 &-0.051 &-0.05 &1.960 &1.956 &0.004 &0.004 &2.578 &2.644 &-0.066 &-0.066 \\
TPSS &1.019 &1.059 &-0.040 &-0.04 &1.907 &1.898 &0.009 &0.009 &2.511 &2.567 &-0.056 &-0.056 \\
\hline
\multicolumn{13}{|l|}{Global Hybrid Generalized Gradient Approximation (GH GGA)} \\
\hline
B3LYP &0.950 &1.070 &-0.120 &-0.11 &1.936 &2.004 &-0.068 &-0.067 &2.592 &2.746 &-0.154 &-0.154 \\
PBE0 &0.893 &1.061 &-0.168 &-0.169 &1.922 &2.035 &-0.114 &-0.113 &2.593 &2.816 &-0.223 &-0.224 \\
BHHLYP &0.634 &1.067 &-0.433 &-0.434 &1.847 &2.208 &-0.361 &-- &2.529 &3.110 &-0.581 &-0.580 \\
\hline
\multicolumn{13}{|l|}{Global Hybrid Meta-Generalized Gradient Approximation (GH meta-GGA)} \\
\hline
TPSSH &0.956 &1.054 &-0.099 &-0.099 &1.913 &1.955 &-0.043 &-0.043 &2.543 &2.672 &-0.129 &-0.131 \\
M08-HX &1.159 &1.294 &-0.135 &-0.135 &2.216 &2.330 &-0.114 &- &2.931 &3.138 &-0.207 &-0.208 \\
M06-2X &1.106 &1.261 &-0.155 &-0.155 &2.200 &2.339 &-0.139 &- &2.961 &3.182 &-0.221 &-0.221 \\
\hline
\multicolumn{13}{|l|}{Range-Separated Hybrid Generalized Gradient Approximation (RSH GGA)} \\
\hline
$\omega$B97X &0.784 &1.204 &-0.421 &-0.421 &2.026 &2.381 &-0.354 &-0.354 &2.787 &3.278 &-0.491 &-0.492 \\
$\omega$B97X-V &0.841 &1.230 &-0.390 &-0.39 &2.071 &2.407 &-0.336 &-0.337 &2.845 &3.303 &-0.457 &-0.458 \\
$\omega$B97X-D3 &0.844 &1.194 &-0.350 &-0.35 &2.037 &2.328 &-0.291 &-0.292 &2.784 &3.197 &-0.412 &-0.412 \\
CAMB3LYP &0.819 &1.126 &-0.307 &-0.308 &1.959 &2.207 &-0.248 &-0.248 &2.667 &3.039 &-0.372 &-0.373 \\
LRC-$\omega$PBEh &0.808 &1.127 &-0.319 &-0.32 &1.953 &2.205 &-0.252 &-0.253 &2.665 &3.038 &-0.373 &-0.373 \\
$\omega$B97M - V &0.968 &1.256 &-0.288 &-0.289 &2.142 &2.396 &-0.254 &-0.254 &2.912 &3.266 &-0.353 &-0.352 \\
\hline
\multicolumn{13}{|l|}{Double Hybrid Generalized Gradient Approximation (DH GGA)} \\
\hline
$\omega$B97x-2(TQZ) &1.478 &1.160 &0.318 &0.294 &2.414 &2.203 &0.211 &- &3.533 &3.009 &0.524 &- \\
$\omega$B97x-2(LP) &1.557 &1.187 &0.370 &0.372 &2.418 &2.185 &0.233 &- &3.624 &2.982 &0.642 &- \\
XYGJOS &1.106 &1.189 &-0.083 &-0.085 &2.264 &2.218 &0.046 &- &2.956 &2.989 &-0.033 &- \\
PBE02 &1.610 &1.138 &0.472 &0.433 &2.397 &2.146 &0.251 &- &3.665 &2.952 &0.713 &- \\
B2PLYP &1.277 &1.092 &0.185 &0.18 &2.256 &2.038 &0.218 &- &3.064 &2.784 &0.280 &- \\
PBE-QIDH &1.244 &1.102 &0.142 &0.125 &2.256 &2.145 &0.111 &- &3.217 &2.977 &0.240 &- \\
B2GP-PLYP &1.386 &1.101 &0.285 &0.273 &2.252 &2.121 &0.131 &- &3.287 &2.856 &0.431 &- \\
\hline
TD-$\omega$PBEPP86 & 1.110 & 1.172 & -0.062 & &  2.310 & 2.242 & 0.068 &  & 2.863 & 2.995 & -0.132 & \\
NEVPT2(12,9) & 0.964 & 1.055 & -0.091 & &  2.246 & 2.231 & 0.015 & &  2.59 & 2.77 & -0.18 & \\
EOM-CCSD & 1.068 & 1.146 & -0.077 & & 2.382 & 2.213 & 0.169 & & 2.916 & 3.066 & -0.150 & \\
FICMRCI(12,9)\cite{chanda2023nature} & 1.022 & 1.099 & -0.077 & & 2.249 & 2.235 & 0.014 & &  2.829 & 2.994 & -0.165 & \\

\hline
\end{tabular}
\end{table}

\subsection{Evaluation of $\Delta$SCF performance}
The MOM and SGM techniques allow us to compute the S$_1$ excited state within the ground state DFT framework. This allows the capture of correlation at the same level for various states without employing further approximations on the excitation ansatz. This is not to say that a fully balanced description can be obtained for all states as the same XC-functional may have different extent of accuracy for different electronic states, but it is a viable option in the DFT framework. We have observed a correct qualitative prediction of inversion or not as well as the relative size of the STGs for the different molecules for non-hybrid functionals with this technique. The numbers are reported in Table~IV. Moreover, the results are quite encouraging because of the inability of LR-TDDFT with standard XC functionals to compute inverted singlet-triplet gaps. The DH DFA's discussed in Sec.~IV.A are more accurate but also more costly. The use of a variational
DFT framework with non-hybrid functionals would allow computations on far larger molecule of interest as well as their n-mers. The MOM/SGM techniques can capture dynamic spin-polarization through state-specific orbital optimization in the presence of correlation. In the unrestricted framework this is reflected in the consequent spin contamination of the wave function. 
Non-hybrid functionals give the best results in this $\Delta$SCF framework. The $\Delta$E$_{ST}$ starts to become too negative from GH GGA onward, which indicates that HF exchange widens the singlet-triplet gap even in this case, although in the opposite direction. This can be traced to greater spin contamination as we shall discuss in Sec.~IV.C. 

The $\Delta$SCF-DFT contrary to expectation, gives lower errors for $\Delta$E$_{ST}$ than for the excitation energies relying on error cancellation across all functionals except double hybrids. MOM results are benchmarked with ICMRCI (12,9) in Fig.~5 in terms of MAE for a hierarchy of XC functionals. Interestingly, TPSSH which only contains 10\% HF exchange, predicts the best $\Delta$E$_{ST}$ results for all molecules among the hybrid functionals. B3LYP with 20\% HFX comes a close second. XC functionals with high HFX but empirical parameterization do somewhat better (M08-HX, M06-2X) but it is difficult to attribute it to specific features. Recall that it underestimates T$_1-$S$_0$ with TDDFT. 
The over-inversion of the $\Delta$E$_{ST}$ gap in the case of RSH GGA may be due to the presence of (100\%) non-local HF exchange at long-range along with variants of short-range exchange-correlation functionals that increase the separation between $S_1$ and $T_1$ possibly due to the empirical choice of $\omega$, the range-separation parameter, leading to self interaction errors. We have not attempted to tune the $\omega$ parameter. In the case of DH GGA, the $\Delta$E$_{ST}$ gaps are large positive and completely wrong. The errors come from very large S$_1-$S$_0$ gaps. This possibly occurs due to the MP2 correlation energy of S$_1$ being estimated using a formula where the orbitals are assumed to come from S$_0$ or some such incompatibility of the $\Delta$SCF and DH functionality. This needs to be further investigated from the program's implementation point of view and is beyond the scope of this work. Furthermore, it can be observed that the SGM technique fails to optimize the excited-state $S_1$ and collapses to a lower energy state $S_0$ for Azine-4N and azine-7N in the case of DH GGAs. Signed errors for our primary and secondary test set are available in the Table~S3 and S4 in the supplementary material.

\begin{figure}
\centering
\includegraphics[width=1.0\linewidth] {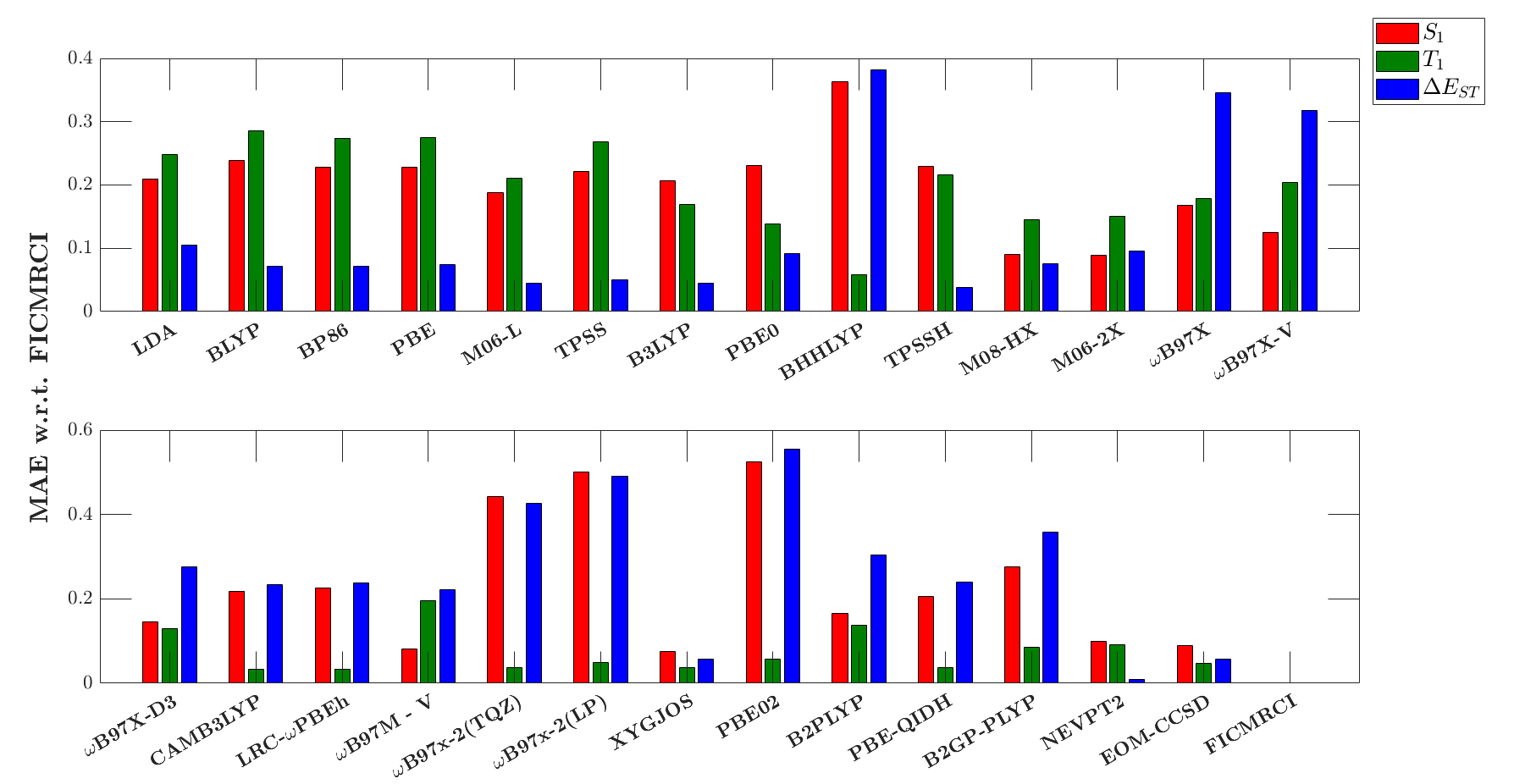}
\caption{MAE in S$_1$, T$_1$, and $\Delta$E$_{ST}$ in eV between MOM with different methods and the ICMRCI (12,9)}
\label{mom_plot} 
\end{figure}

\begin{sidewaystable} 
\caption{Vertical excitation energies and associated $\Delta$E$_{ST}$ energy difference (all in eV) calculated with MOM for 6 other templates. Geometries are optimized at B3LYP/def2-TZVP level}\label{tab: }
\centering
\scriptsize
\begin{tabular}{ccccccccccccccccccc}
\hline	
Molecules & & azine-3N & & & azine-4N(2) & & & azine-5N(1) & & & azine-5N(2) & & & azine-6N(1) & & & azine-6N(2) & \\
\hline
Method & S$_1$ & T$_1$ & $\Delta E_{ST}$ & S$_1$ & T$_1$ & $\Delta E_{ST}$ & S$_1$ & T$_1$ & $\Delta E_{ST}$ & S$_1$ & T$_1$ & $\Delta E_{ST}$ & S$_1$ & T$_1$ & $\Delta E_{ST}$ & S$_1$ & T$_1$ & $\Delta E_{ST}$ \\
\hline
LDA & 1.599 & 1.569 & 0.03 & 1.858 & 1.822 & 0.036 & 2.086 & 2.055 & 0.031 & 2.058 & 2.027 & 0.031 & 2.328 & 2.302 & 0.026 & 1.952 & 1.9 & 0.052 \\
BLYP & 1.512 & 1.5159 & -0.0039 & 1.773 & 1.767 & 0.006 & 1.997 & 1.999 & -0.002 & 1.967 & 1.969 & -0.002 & 2.235 & 2.243 & -0.008 & 1.862 & 1.831 & 0.031 \\
BP86 & 1.519 & 1.523 & -0.004 & 1.783 & 1.778 & 0.005 & 2.011 & 2.013 & -0.002 & 1.981 & 1.983 & -0.002 & 2.252 & 2.262 & -0.01 & 1.877 & 1.843 & 0.034 \\
PBE & 1.5219 & 1.5224 & -0.0005 & 1.785 & 1.776 & 0.009 & 2.013 & 2.012 & 0.001 & 1.982 & 1.981 & 0.001 & 2.254 & 2.26 & -0.006 & 1.880 & 1.843 & 0.037 \\
M06-L & 1.532 & 1.565 & -0.033 & 1.814 & 1.833 & -0.019 & 2.05 & 2.082 & -0.032 & 2.019 & 2.046 & -0.027 & 2.3 & 2.342 & -0.042 & 1.907 & 1.885 & 0.022 \\
TPSS & 1.492 & 1.519 & -0.027 & 1.766 & 1.781 & -0.015 & 1.998 & 2.023 & -0.025 & 1.964 & 1.989 & -0.025 & 2.241 & 2.276 & -0.035 & 1.857 & 1.833 & 0.024 \\
B3LYP & 1.477 & 1.571 & -0.094 & 1.781 & 1.859 & -0.078 & 2.032 & 2.128 & -0.096 & 1.997 & 2.085 & -0.088 & 2.297 & 2.414 & -0.117 & 1.869 & 1.883 & -0.014 \\
PBE0 & 1.437 & 1.58 & -0.143 & 1.756 & 1.880 & -0.124 & 2.015 & 2.163 & -0.148 & 1.978 & 2.113 & -0.135 & 2.289 & 2.462 & -0.173 & 1.849 & 1.889 & -0.040 \\
BHHLYP & 1.251 & 1.626 & -0.375 & 1.624 & 1.966 & -0.342 & 1.906 & 2.291 & -0.385 & 1.860 & 2.209 & -0.349 & 2.204 & 2.639 & -0.435 & 1.694 & 1.869 & -0.175 \\
TPSSH & 1.462 & 1.543 & -0.081 & 1.758 & 1.822 & -0.064 & 2.003 & 2.083 & -0.08 & 1.966 & 2.041 & -0.075 & 2.259 & 2.355 & -0.096 & 1.849 & 1.855 & -0.006 \\
M08-HX & 1.717 & 1.828 & -0.111 & 2.045 & 2.140 & -0.095 & 2.313 & 2.431 & -0.118 & 2.282 & 2.385 & -0.103 & 2.606 & 2.756 & -0.15 & 2.138 & 3.096 & -0.958 \\
M06-2X & 1.69 & 1.808 & -0.118 & 2.032 & 2.130 & -0.098 & 2.316 & 2.433 & -0.117 & 2.282 & 2.384 & -0.102 & 2.62 & 2.776 & -0.156 & 2.126 & 2.107 & 0.019 \\
wB97X & 1.4356 & 1.761 & -0.3254 & 1.817 & 2.095 & -0.278 & 2.113 & 2.412 & -0.299 & 2.077 & 2.343 & -0.266 & 2.435 & 2.774 & -0.339 & 1.914 & 1.987 & -0.073 \\
wB97X-V & 1.492 & 1.788 & -0.296 & 1.871 & 2.122 & -0.251 & 2.171 & 2.442 & -0.271 & 2.135 & 2.373 & -0.238 & 2.493 & 2.806 & -0.313 & 1.967 & 2.019 & -0.052 \\
wB97X-D3 & 1.474 & 1.748 & -0.274 & 1.841 & 2.076 & -0.235 & 2.131 & 2.388 & -0.257 & 2.095 & 2.321 & -0.226 & 2.442 & 2.738 & -0.296 & 1.822 & 1.992 & -0.170 \\
CAMB3LYP & 1.416 & 1.669 & -0.253 & 1.768 & 1.989 & -0.221 & 2.045 & 2.293 & -0.248 & 2.008 & 2.229 & -0.221 & 2.341 & 2.625 & -0.284 & 1.861 & 1.935 & -0.074 \\
LRC-wPBEh & 1.412 & 1.672 & -0.26 & 1.765 & 1.991 & -0.226 & 2.042 & 2.295 & -0.253 & 2.006 & 2.23 & -0.224 & 2.34 & 2.624 & -0.284 & 1.864 & 1.938 & -0.074 \\
wB97M - V & 1.594 & 1.811 & -0.217 & 1.958 & 2.139 & -0.181 & 2.25 & 2.451 & -0.201 & 2.218 & 2.39 & -0.172 & 2.566 & 2.808 & -0.242 & 2.052 & 2.066 & -0.014 \\
\hline
wB97x-2(TQZ) & 2.1444 & 1.974 & 0.1704 & 2.273 & 2.391 & -0.118 & 2.833 & 2.571 & 0.262 & 2.802 & 2.535 & 0.267 & 3.12 & 2.948 & 0.172 & 2.705 & 2.253 & 0.452 \\
wB97x-2(LP) & 2.227 & 2.008 & 0.219 & 2.391 & 2.273 & 0.118 & 2.922 & 2.593 & 0.329 & 2.891 & 2.568 & 0.323 & 3.181 & 2.982 & 0.199 & 2.805 & 2.28 & 0.525 \\
XYGJOS & 1.693 & 1.736 & -0.043 & 2.044 & 2.058 & -0.014 & 2.326 & 2.347 & -0.021 & 2.276 & 2.294 & -0.018 & 2.624 & 2.643 & -0.019 & 2.134 & 2.098 & 0.036 \\
PBE02 & 2.276 & 2.019 & 0.257 & 2.370 & 2.315 & 0.055 & 2.97 & 2.608 & 0.362 & 2.936 & 2.585 & 0.351 & 3.182 & 3.004 & 0.178 & 2.860 & 2.302 & 0.558 \\
B2PLYP & 1.843 & 1.695 & 0.148 & 2.152 & 2.005 & 0.147 & 2.443 & 2.301 & 0.142 & 2.410 & 2.264 & 0.146 & 2.73 & 2.615 & 0.115 & 2.295 & 2.084 & 0.211 \\
PBE-QIDH & 1.8816 & 1.824 & 0.0576 & 2.172 & 2.145 & 0.027 & 2.549 & 2.46 & 0.089 & 2.513 & 2.408 & 0.105 & 2.839 & 2.82 & 0.019 & 2.400 & 2.138 & 0.262 \\
B2GP-PLYP & 1.994 & 1.804 & 0.19 & 2.283 & 2.119 & 0.164 & 2.633 & 2.425 & 0.208 & 2.601 & 2.383 & 0.218 & 2.921 & 2.769 & 0.152 & 2.495 & 2.163 & 0.332 \\
\hline
TD-wPBEPP86 & 1.698 & 1.712 & -0.014 & 2.058 & 2.107 & -0.049 & 2.316 & 2.377 & -0.061 & 2.267 & 2.281 & -0.014 & 2.583 & 2.685 & -0.102 & 2.171 & 2.168 & 0.003 \\
NEVPT2(12,9) & 1.495 & 1.611 & -0.116 & 2.013 & 2.012 & 0.001 & 2.543 & 2.511 & 0.032 & 2.137 & 2.178 & -0.041 & 2.508 & 2.605 & -0.097 & 2.374 & 2.016 & 0.358 \\
EOM-CCSD & 1.7033 & 1.7274 & -0.0241 & 2.054 & 2.0421 & 0.0119 & 2.245 & 2.2712 & -0.0258 & 2.2454 & 2.2712 & -0.0258 & 2.5717 & 2.6349 & -0.0632 & 2.155 & 2.0765 & 0.0785 \\
\hline
\end{tabular}
\end{sidewaystable}

The errors of the individual molecules in S$_1$, T$_1$ and $\Delta$E$_{ST}$ are similar up to GH-GGA level but start to vary significantly for RSH and DH functionals. In particular, azine-4N gives much larger errors with only XYGJOS (DH DFA) doing reasonably well. Nevertheless, the error in $\Delta$E$_{ST}$  for non-hybrid exchange-correlation functionals up to the meta-GGA level is minimal. The $\Delta$E$_{ST}$ gap begins to diverge after GH GGA because the error in $S_1$ grows, while $T_1$ shows no discernible change in the case of azine-1N. For azine-7N, from GH-GGA to RSH GGA level, $S_1$ value has hugely deviated in comparison with $T_1$ and $\Delta$E$_{ST}$ and for DH GGA, all individual $S_1$, $T_1$, and $\Delta$E$_{ST}$ have significantly deviated from ICMRCI (12,9) results. Considering individual errors in S$_1-$S$_0$, T$_1-$S$_0$ and $\Delta$E$_{ST}$ for MOM the XC functionals with MAE$<$0.15 eV are XYGJOS $<$ M08-HX $<$ M06-2X $<$ B3LYP $<$ M06-L $<$ PBE0 with XYGJOS being the most accurate and stable among various molecules with least MAE of ~0.05 eV. Apart from that PBE-QIDH $<$ TPSSH $<$ CAMB3LYP $<$ LRC-$\omega$PBEh $<$ $\omega$B97M-V $<$ TPSS $<$ $\omega$B97X-D3 $<$ LDA $<$ BP86 $<$ PBE $<$ BLYP are the ones with MAE$<$0.2 in ascending order of MAE. However, if we take a closer look at the signed error in Table~S2 of the supplementary material and relative errors of S$_1$ and T$_1$and across all the molecules, one realizes that the non-hybrid functionals are far more equitable in their errors across the three molecules and also correctly predicts the relative degree of inversion. The absolute error in the predicted S$_1$ and T$_1$ excitation energies are somewhat larger leading to a larger MAE. A systematic improvement is also noted from LDA$\to$ GGA$\to$ meta-GGA which is absent for the hybrid DFAs. Based on these arguments and also the lower computational cost, we recommend TPSS and M06-L (meta-GGA). For larger systems BLYP, PBE and BP86 (GGA) will also suffice. If hybrid functionals are necessary due to other physics in other parts of the candidate molecules, TPSSH may be preferred for the $\Delta$SCF computations over the other hybrid functionals. The performance of XYGJOS is noteworthy. XYGJOS is a XYG3 type of DH functional which is based on the adiabatic connection and the Gorling-Levy perturbation theory of second order. It uses the opposite-spin (OS) ansatz combined with locality of electron correlation which has also been found to improve the results of the LR-TDDFT in combination with other functionals. Unfortunately an implementation of XYGJOS in the LR-TDDFT framework was not available to us. XYGJ-OS involves a doubly hybrid density functional (DHDF), containing both a nonlocal orbital-dependent component in the exchange term (HF-like exchange), and also information about the unoccupied KS orbitals in the electron correlation part (PT2, perturbation theory up to second order) using the opposite-spin (OS) ansatz to include the locality of electron correlation. The proposition of a unique OS ansatz for DHDF, yields a balanced description of nonlocal correlation effects while considerably reducing computational time. The OS ansatz is motivated by the observation that the most important electron correlation effects involve correlations of the OS electrons in the same orbital. It should be emphasized that as the same-spin (SS) correlation is very important in accurate description of open-shell systems and magnetic properties, such contributions cannot be simply neglected. In XYGJ-OS, the SS correlation effects are included within the standard DFA. Fully optimized B3LYP orbitals are used to generate the density and to calculate each term in PT2. But here only the energetically most important double excitation PT2 terms are taken explicitly using orbitals generated from a conventional DFA and the present OS ansatz further reduces the computational cost by only calculating the most important electron correlation effects contributed by the OS electrons in the same orbital. B3LYP is anyhow performing well with MOM for all cases with a MAE of~$<$0.15. Including B3LYP orbitals to generate opposite spin (OS) correlation effects is similar to inclusion of spin polarization which leads to higher accuracy. Similar situation is observed in Ref\cite{chanda2023nature} where OO-MP2 and SOS-OO-MP2 were better to model T$_1-$S$_0$ excitation energies compared to MP2 due to a balanced description of orbital optimization and spin polarization.        


\begin{table}[h]
\centering
\caption{S$_1$, T$_1$ and $\Delta$E$_{ST}$ along with $<S^2>$ value as a function of fraction of HF exchange for MOM-B3LYP/def2-TZVP}\label{tab: }
\scriptsize
\begin{tabular}{cccccc}
\hline
& & & azine-1N & & \\
\hline
HF Exchange & S$_1$ & T$_1$ & $\Delta$E$_{ST}$(eV) & $<S^2>_{S_1}$ & $<S^2>_{T_1}$ \\
\hline
0 & 1.0721134 & 1.08054881 & -0.00843541 & 1.00991 & 2.00172 \\
0.1 & 1.02394993 & 1.07592294 & -0.05197301 & 1.05364 & 2.0066 \\
0.2 & 0.959106117 & 1.072712042 & -0.113605925 & 1.09013 & 2.0106 \\
0.4 & 0.77632983 & 1.07184129 & -0.29551146 & 1.19696 & 2.0225 \\
0.6 & 0.51918588 & 1.07918826 & -0.56000238 & 1.34808 & 2.04175 \\
0.8 & 0.18911645 & 1.09633119 & -0.90721474 & 1.5336 & 2.072 \\
1 & -0.04108861 & 1.11075302 & -1.15184163 & 1.65447 & 2.09834 \\
\hline
& & & azine-4N & & \\
\hline
HF Exchange & S$_1$ & T$_1$ & $\Delta$E$_{ST}$(eV) & $<S^2>_{S_1}$ & $<S^2>_{T_1}$ \\
\hline
0 & 1.911246218 & 1.885341346 & 0.025904872 & 1.01911 & 2.00327 \\
0.1 & 1.931028615 & 1.942729345 & -0.01170073 & 1.0396 & 2.0059 \\
0.2 & 1.936035439 & 2.00409015 & -0.068054711 & 1.06812 & 2.00968 \\
0.4 & 1.903872037 & 2.14096148 & -0.237089443 & 1.1448 & 2.02132 \\
0.6 & 1.82177645 & 2.29905739 & -0.47728094 & 1.22416 & 2.04039 \\
0.8 & 1.71266034 & 2.47864999 & -0.76598965 & 1.26428 & 2.07102 \\
1 & 1.647598839 & 2.597344372 & -0.949745533 & 1.2629 & 2.09809 \\
\hline
& & & azine-7N & & \\
\hline
HF Exchange & S$_1$ & T$_1$ & $\Delta$E$_{ST}$(eV) & $<S^2>_{S_1}$ & $<S^2>_{T_1}$ \\
\hline
0 & 2.51266374 & 2.53116722 & -0.01850348 & 1.02824 & 2.003 \\
0.1 & 2.562296604 & 2.63565746 & -0.073360856 & 1.05678 & 2.00586 \\
0.2 & 2.592528025 & 2.74613412 & -0.153606095 & 1.09783 & 2.01016 \\
0.4 & 2.5932083 & 2.98858413 & -0.39537583 & 1.22164 & 2.024 \\
0.6 & 2.50667732 & 3.26205468 & -0.75537736 & 1.40061 & 2.0471 \\
0.8 & 2.33524802 & 3.56899476 & -1.23374674 & 1.62171 & 2.08367 \\
1 & 2.195900489 & 3.76981194 & -1.573911451 & 1.765705 & 2.11474 \\
\hline
\end{tabular}
\end{table}

\begin{figure}[h]
    \centering
    \includegraphics[width=1.0\linewidth]{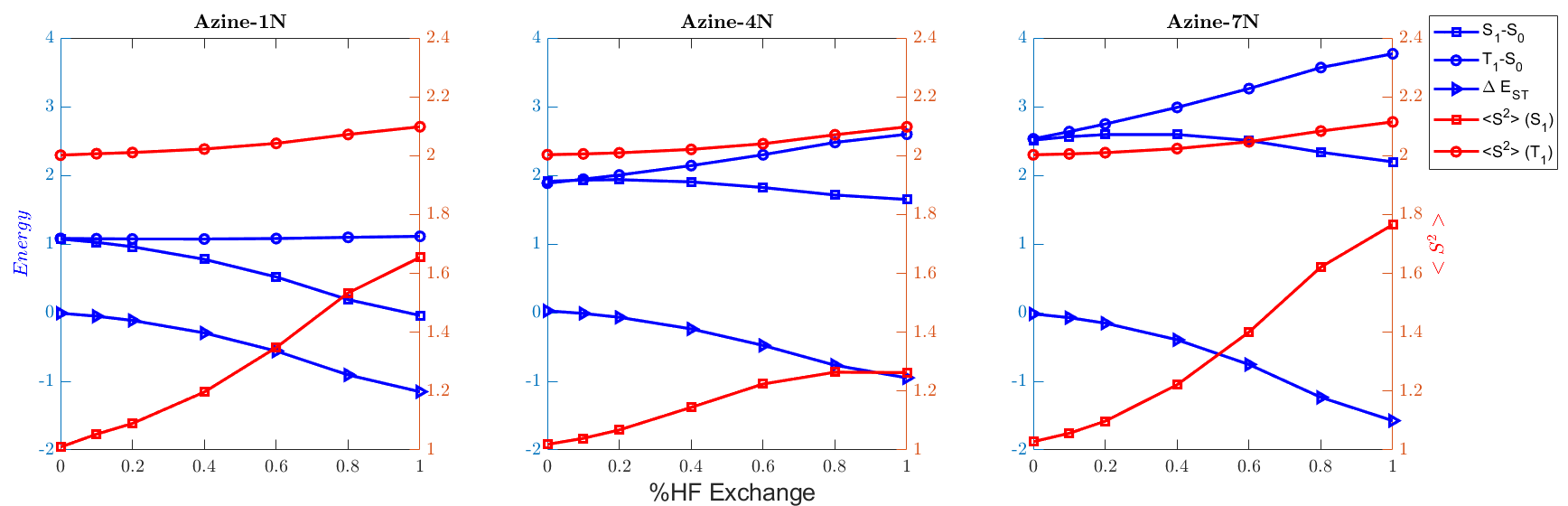}
	\caption{S$_1-$S$_0$, T$_1-$S$_0$ and $\Delta$E$_{ST}$ values with $<S^2>$ as a function of HF exchange for MOM-B3LYP/def2-TZVP calculation.}
    \label{fig:enter-label}
\end{figure}

It is clear from the previous discussion for that the estimated $\Delta$E$_{ST}$ gaps become more negative from GH-GGA onward with the amount of HF exchange in the exchange-correlation functional
Table VI and Fig.~6 show that the computed singlet-triplet gaps from MOM-B3LYP/def2-TZVP calculations, rise in proportion to the degree of HF exchange in the XC functional. With an increase in fraction of HF exchange for MOM-B3LYP, the gap between $S_1$ and $S_0$ shrinks more quickly while the gap between $T_1$ and $S_0$ remains constant or grows a little, which causes the $\Delta$E$_{ST}$ gaps to rise leading to large negative ST gaps. This is in contrary to TDDFT-B2PLYP (as shown in Table~III) where we found that the \%HFX eventually closing the ST gaps leading to a positive $\Delta$E$_{ST}$. These observations merit further analysis and in the next sections we study the role of HF exchange and its correlation with spin-contamination of the open-shell states when using the $\Delta$SCF-DFT procedures.

\begin{table}[!htp]\centering
\caption{$<S^2>$ value of S$_1$ and T$_1$ states for azine-1N using MOM-DFT/def2-TZVP calculations for several XC functionals.}\label{tab:dft_s2}
\scriptsize
\begin{tabular}{|c|cc|cc|cc|}
\hline
Molecules & azine-1N & & azine-4N &  & azine-7N & \\
\hline
Method &S$_1$ &T$_1$ &S$_1$ &T$_1$ &S$_1$ &T$_1$ \\
\hline
LDA &1.013 &2.002  &1.010 &2.002  & 1.014 &2.002 \\
BLYP &1.030 &2.004  &1.021 &2.003  &1.030 &2.003 \\
PBE &1.033 &2.004  &1.022 &2.004  &1.032 &2.003 \\
M06-L &1.070 &2.010  &1.048 &2.009  &1.068 &2.009 \\
TPSS &1.055 &2.007  &1.036 &2.006  &1.052 &2.005 \\
B3LYP &1.090 &2.011  &1.068 &2.010  &1.098 &2.010 \\
PBE0 &1.129 &2.015  &1.094 &2.013  &1.137 &2.013 \\
BHHLYP &1.277 &2.032  &1.191 &2.031  &1.316 &2.037 \\
TPSSH &1.091 &2.011  &1.063 &2.009  &1.091 &2.008 \\
M08-HX & 1.094 & 0.013  & 1.082 & 0.012 & 1.113 & 0.013 \\
M06-2X &1.094 &2.012  &1.086 &2.013  &1.113 &2.014 \\
$\omega$B97X-V &1.211 &2.024  &1.165 &2.021  &1.212 &2.020 \\
$\omega$B97X-D3 &1.203 &2.024  &1.153 &2.020  &1.201 &2.020 \\
CAMB3LYP &1.188 &2.021  &1.140 &2.019  &1.192 &2.019 \\
LRC-$\omega$PBEh &1.199 &2.023  &1.145 &2.019  &1.194 &2.017 \\
$\omega$B97M - V &1.160 &2.019  &1.132 &2.017  &1.166 &2.017 \\
$\omega$B97x-2(TQZ) &1.580 &2.083  &1.269 &2.075  &1.638 &2.079 \\
$\omega$B97x-2(LP) &1.625 &2.092  &1.260 &2.084  &1.690 &2.089 \\
XYGJOS &1.090 &2.011 & 1.068 &2.010  &1.098 &2.010 \\
PBE02 &1.676 &2.106 & 1.894 &2.097  &1.756 &2.105 \\
B2PLYP &1.338 &2.041  &1.213 &2.038  &1.381 &2.043 \\
PBE-QIDH &1.536 &2.074  &1.257 &2.067  &1.597 &2.074 \\
\hline
\end{tabular}
\end{table}

\begin{figure}
    \includegraphics[width=1.0\linewidth]{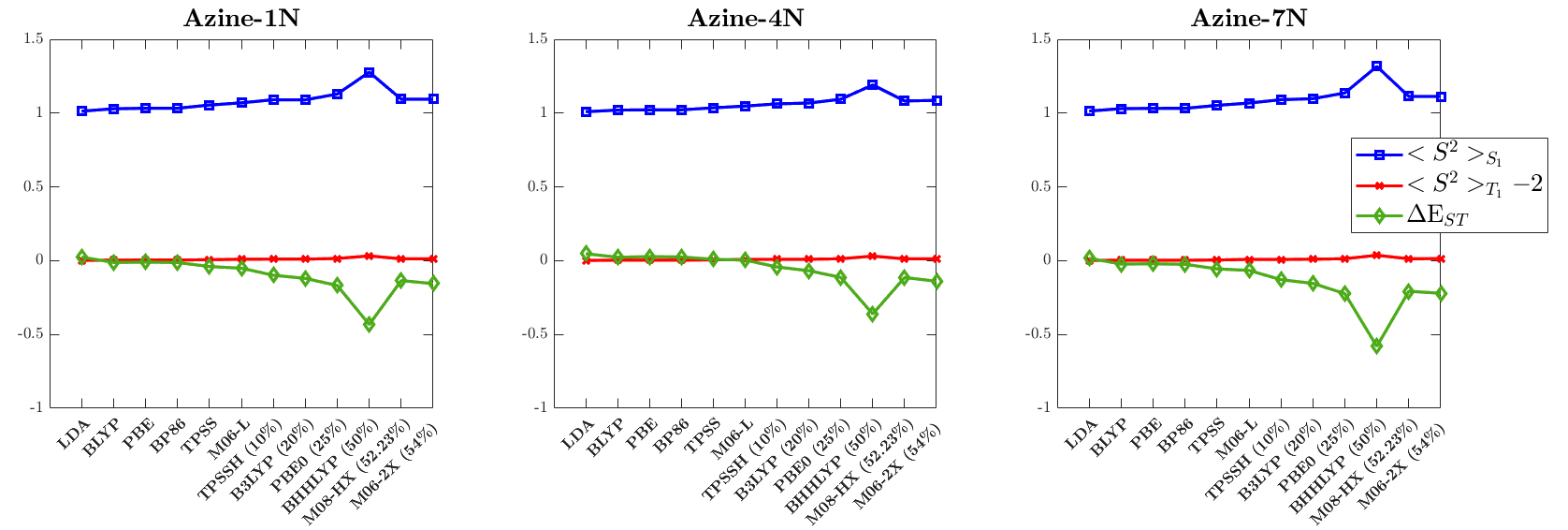}
    \caption{Variation of $<S^2>$ with pure and hybrid DFT functionals. For the hybrid functionals, the percentage of the Hartree–Fock (HF) exchange is given within parentheses}
    \label{fig:dft_hfex}
\end{figure}

\subsection{Spin Contamination in DFT}
Spin contamination is a potential problem whenever unrestricted spin orbitals are variationally optimized for an open-shell state. Furthermore, the problem is more pronounced when a multi-determinant state such as an open-shell singlet state like $S_1$ in our case, is artificially represented as a Kohn-Sham determinant. In the latter case, the flexibility of the KS determinant to include the correlation is enhanced by relaxing the spin symmetry constraint. It is quite remarkable that MOM-DFT does so well in reproducing the excitation energies across several molecules while incurring a huge spin contamination error. Spin contamination of the $S_1$ and $T_1$ states at the MOM-HF/def2-TZVP level ($<S^2>$= 2.0166 and 2.493 for $S_1$ and $T_1$, respectively, for azine-1N) causes erroneous results.
This is mitigated by the MOM-DFT approaches ($<S^2>$= 1.09 and 2.01 for $S_1$ and $T_1$, respectively, for azine-1N at MOM-B3LYP/def2-TZVP), which include more correlation. However, the singlet state ($S_1$) is still highly spin-contaminated due to a strong orbital mixing of $S_0$ and $T_1$ states whereas contamination is almost negligible in the triplet state. Table~VII shows that spin contamination in the singlet state is found to be between 1.0 and 1.7 for all functionals (the ideal value being 0) and near the ideal 2.0 for the triplet state. The LDA functional has the least amount of spin contamination in the singlet state, while PBE0-2 has the most. It indicates that pure DFT functionals (LDA, BLYP, TPSS) are marginally more effective than hybrid functionals (GH GGA, GH meta GGA) at minimizing spin contamination, and it is illustrated in Fig.~7 that with a greater percentage mixing of HF exchange in hybrid DFTs, the KS determinant (i.e. low lying excited state $S_1$ and $T_1$) will be more spin-contaminated and that the $\Delta$E$_{ST}$ perfectly correlates with the $<S^2>$ of S$_1$. M08-HX and M06-2X having greater \%HFX still give lower ST gaps due to less spin contamination. Sufficient correlation must be allowed to modulate the extent of spin contamination in order to be able to mimic the dynamic spin polarization in a single determinant framework. In other words, by introducing spin-breaking in the KS determinant, we mimic part of the effect of dynamic spin-polarization\cite{drwal2023} in a somewhat uncontrolled manner but guided by the extent of correlation captured by the XC functional 
similar to that observed in OO-MP2 for the T$_1$ state\cite{chanda2023nature}. In fact, one can return to positive ST gaps\cite{ye2020arxiv} by eliminating the spin-contamination by using, for instance, the ROKS formalism\cite{hait2016}, which uses a spin pure reference determinant to produce excited SCF states that are orthogonal (e.g., UB3LYP gives a ST gap -0.12 eV with $<S^2>$ value 1.09 and 2.011 for S$_1$, T$_1$, while ROKS-B3LYP gives a ST gap ~0.13 eV). The $<S^2>$ value (spin-contamination) is the only variation in this case, which amply illustrates its significance in the inversion of the ST gap during state-specific orbital optimisation. Furthermore, it is important to highlight that KS orbitals have recently been used for post-SCF computations based on the argument that MOM-DFT methods are significantly less spin-contaminated than the UHF method\cite{rettig2020,fang2016,beran2003,MALLICK2021113326}. Other approaches that incorporate orbital relaxation effects into variational excited state solutions inside DFT exhibit comparable performance when compared to the MOM approach\cite{hait2016,hait2020,hait2021}. The direct correlation between the \%HFX and spin contamination of the wave function is visible in Table~VI and Fig.~6, where we found that the for a fixed XC functional (B3LYP), if we tune the fraction of \%HFX (also adjusting DFT exchange to make the total exchange unity), it will increase the spin-contamination of the S$_1$ wave-function leading to larger $<S^2>$ value. For T$_1$, the $<S^2>$ value apparently remain the same with increasing \%HFX. Eventually increasing negative $\Delta$E$_{ST}$ with increasing \%HFX proves its direct connection with the spin-polarization and hence it is a crucial factor for ST inversion in this type of molecules.

\begingroup
\begin{table}
\centering
\caption{Kif integral and ROKS ST gap}\label{tab:}
\begin{tabular}{c|ccc|ccc|ccc}
\hline
	& & Azine-1N & & & Azine-4N & & & Azine-7N &  \\
\hline
& K$_{if}$ & ROKS & MOM & K$_{if}$ & ROKS & MOM & K$_{if}$ & ROKS & MOM \\
\hline
B3LYP & 0.1364 & 0.1123 & -0.1100 & 0.1990 & 0.1274 & -0.068 & 0.1523 & 0.1198 & -0.154 \\
TPSSH & 0.1414 & 0.1319 & -0.099 & 0.2051 & 0.1439 & -0.0430 & 0.1592 & 0.1336 & -0.129 \\
PBE0 & 0.1302 & 0.1346 & -0.1690 & 0.1869 & 0.1493 & -0.1140 & 0.1389 & 0.1383 & -0.223 \\
M06 & 0.1350 & 0.1229 & -0.1350 & 0.2041 & 0.1344 & -0.1140 & 0.1476 & 0.1238 & -0.207 \\
M06-2X & 0.1306 & 0.0909 & -0.1550 & 0.1867 & 0.1111 & -0.1390 & 0.1397 & 0.1003 & -0.221 \\
TPSS & 0.1448 & 0.1232 & -0.0400 & 0.2111 & 0.1045 & 0.0090 & 0.1659 & 0.1336 & -0.056 \\
\hline
\end{tabular} 
\end{table}
\endgroup

\begin{figure}
    \centering
    \includegraphics[width=1.0\linewidth]{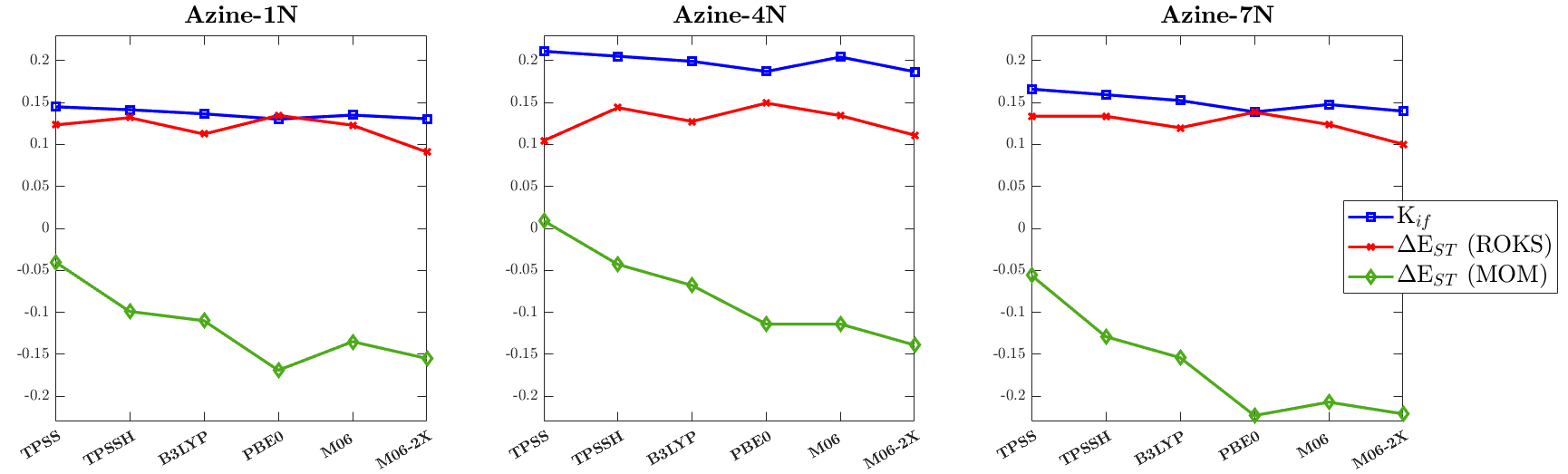}
	\caption{K$_{if}$ integral with ROKS and MOM $\Delta$E$_{ST}$ energy values in eV for different methods for azine-1N, azine-4N and azine-7N}
\label{tddft_fig}   
\end{figure}


Recently Omar et. al.\cite{omar2023} developed a virtual screening strategy in which the molecules to be examined are pre-selected using progressively more complex filters, such as the approximation of CASSCF calculations and the evaluation of the HOMO-LUMO exchange integral (K$_{if}$). By taking a large test set of molecules with properties similar to known INVEST dyes, thay have set K$_{if}$ $<$ 0.4 eV as the upper boundary for this quantity in the screening. This increases the likelihood of successfully identifying new INVEST molecules by searching for low K$_{if}$=K$_{HL}$. A tiny exchange coupling between the unpaired electrons which signifies their CT characters is not an exclusive condition for ST inversion and the presence of an extra mechanism stabilising the singlet state over the triplet state is required for a singlet–triplet state inversion to occur. Multi-reference character of the excited states helps the situation by increasing the total correlation. We have also evaluated the K$_{if}$ integrals for a subset of XC functionals from the set for MOM calculations. The lowest excited singlet-triplet gap is equal to 2K$_{if}$ at the CIS level of (uncorrelated) theory. This result is same as the `two state model' as discussed in Ref\cite{desilva2019a}. But according to the Becke's exciton model\cite{becke2018} for a correlated SCF (DFT), it equals to the integral K$_{if}$ itself through a virial exciton theorem, where i and f are two unpaired open-shell electrons. The lowest excited singlet and triplet states in our closed-shell systems are dominated by singly excited configurations that means a predominantly HOMO $\to$ LUMO transition (CT). To understand the effect of double excitations on the value and sign of the ST gap, a minimal three-state model is considered in ref\cite{desilva2019b} where the basis states are two spin-mixed CT configurations {$|CT_1⟩, |CT_2⟩$} ($\phi_a \to \phi_b$ transitions) and a doubly excited configuration $|D⟩$ ($(\phi_a)^2 \to (\phi_b)^2$ transition). The coupling between $|CT⟩$ and $|D⟩$ states is the sum of one- and two-electron integrals $t_{CT,D} = h_{AB} + (\phi_a \phi_b|\phi_b \phi_b)$, known as the hopping integral. Because doubly excited configurations $|\psi_{i,\overline{i}}^{a,\overline{a}}>$ can mix only with CSFs of singlet multiplicity, adding them leads to the reduction of the singlet–triplet gap. If the coupling is strong (large $t_{CT,D}^2$) and/or singly and doubly excited configurations are energetically close (small $E_D$), a substantial admixture of the double excitation in the S$_1$ state can lead to an inverted singlet–triplet gap, which would be a manifestation of the electron correlation effect in the excited state. This is effectively the same mechanism as the kinetic exchange\cite{koch2012}, which is explained as a two-step process where the electrons are swapped by pairing them first in one orbital\cite{difley2008}. The singlet–triplet gap predicted by the kinetic exchange model is given by, 
\begin{equation}
    \Delta E_{ST} = K_{if} - \frac{t^2}{\Delta E}
\end{equation}
where $K$ is the direct exchange coupling, $t$ is the hopping integral between involved orbitals, and $\Delta E$ is the energy difference between the paired and unpaired states. The second term, which is particularly related to doubly excited configurations, is a perturbative estimate of the singlet stabilisation owing to dynamic spin polarisation\cite{drwal2023,bedogni2024}, closely analogous to the kinetic exchange energy stabilisation in Heisenberg antiferromagnets\cite{ANDERSON1963}. In these heterocyclic fused rings, the exchange energy - which in most systems causes triplet states to lie lower than singlet states is insignificant, and it is possible that the normally insignificant impact of dynamical spin polarisation will come into play and flip the ST gap. In particular, the interaction with the doubly excited state - where both electrons sit on the LUMO - stabilizes the singlet excited state. The corresponding triplet states cannot be stabilised in this way.


In the context of MOM-DFT, the effect of the major contributions of the dynamic spin polarization is captured through the spin-contamination of the spin-broken determinant. In our case we got K$_{if}$ values less than the upper boundary of 0.4 eV proposed\cite{omar2023} as these are well known INVEST molecules. These values along with the $\Delta$E$_{ST}$ from MOM and ROKS for a set of functionals are tabulated for azine-1N, azine-4N and azine-7N in Table~VIII. The ROKS method is difficult to converge for many functionals. It is observed that even where MOM-DFT gives accurate negative ST gaps, ROKS gives small positive numbers. 
In fact using a spin-pure reference, the ST gap is nothing but the HOMO-LUMO exchange energy, designated as the K$_{if}$ integral which are comparable for most of the cases. For a few functionals especially in azine-4N , it deviates a little bacause the ROKS method sometimes converges poorly. A pictorial representation of correlation between these quantities is depicted in Fig.~8. The larger the K$_{if}$ value, less negative or more degenerate is the ST gap. Eventually if the K$_{if}$ crosses a certain threshold, the ST gap becomes positive. For example, the K$_{if}$ value is 0.21 eV for azine-4N which has a positive STG (0.009 eV with TPSS/def2-TZVP). This is less than the 0.4 eV threshold indicating that K$_{if}$ is not the sole criterion as we have strained earlier. For comparison, phenalene ($C_{13}H_{10}$) having $\Delta$E$_{ST}$= 0.38 eV with TPSS/def2-TZVP, has a K$_{if}$ of 1.27 eV. From the Becke's virial exciton model\cite{becke2018}, the K$_{if}$ integrals are found to increase with increasing \%HF exchange\cite{becke2019} in XC functionals and correlates with the trend of ST gaps becoming more positive as we have seen previously for TDDFT. But in case of MOM based DFT methods as we have seen spin-contamination (leading to dynamic spin polarization) also increases with increasing \%HFX. Here, both the terms on the RHS of Eq. (9) increase with \%HFX but the dominant contributions are from the hopping (t) term (which correlates with the spin-contamination) leading to more negative STG with more \%HFX. The positivity or near degeneracy of STG lies in a proper balance of both these terms in Eq. (9) and $\frac{t^2}{\Delta E} > K_{if}$ is the minimum condition for ST inversion as discussed in Ref\cite{desilva2019a}. For excited states with predominant CT character, the hopping integral term is negligible compared to K$_{if}$, hence Eq.~(9) reduces to $\Delta E_{ST} \approx K_{if}$, which is nothing but the Becke's exciton model\cite{becke2018}.

\section{CONCLUSIONS}
In this paper, we benchmark the DFT-based excited state methodologies with Internally Contracted Multireference Configuration Interaction Singles Doubles (ICMRCISD) method for the challenging cases of INVEST molecules with near degenerate S$_1$ and T$_1$. 
The present study confirms that accurate gaps including inversion of singlet-triplet can be obtained using the LR-TDDFT(D) approach with most of the functionals especially those with spin-component scaling. The best performing XC functionals for LR-TDDFT(D) are $\omega$PBEPP86 $<$ SCS-$\omega$B88PP86 $<$ SCS-PBEQIDH $<$ PBE0-2 $<$ B2GP-PLYP $<$ $\omega$B88PP86 ordered in ascending average error. SOS-$\omega$PBEPP86, B2PLYP and PBE-QIDH are also close in accuracy. Our suggestions are based not only on the $\Delta$E$_{ST}$ but also the S$_1-$S$_0$ and T$_1-$S$_0$ gaps which play a crucial role in the fluorescence process. These findings suggest that to obtain inverted singlet-triplet gaps, correlation originating from double excitations must be included with their major effect being the spin polarization due to correlation (also called dynamic spin polarization). 

On the other hand, we have found that the $\Delta$SCF methods (MOM, SGM) are also quite successful in predicting the ST gaps of these molecules albeit with a slightly lower accuracy. The sign and relative ST gaps of various molecules are well predicted with the same non-hybrid XC functional which is a significant advantage for a screening method along with the low computational cost. Due to the spin-breaking of the wave function, these methods are largely spin-contaminated, still the results are quite stable with standard XC functionals. The best performing XC functionals for MOM and SGM based on MAE are XYGJOS $<$ M08-HX $<$ M06-2X $<$ B3LYP $<$ M06-L $<$ PBE0 but our recommendation for screening are TPSS and M06-L which have a low standard deviation and variance across molecules. In case substituents on these templates need hybrid functionals, TPSSH should be the functional of choice. If the molecules are too big, one may use BLYP, PBE, or BP86.

Removing spin-contamination through the ROKS method leads to small positive $\Delta$E$_{ST}$. We may surmise that spin contamination is a crucial factor for the ST inversion in DFT based $\Delta$SCF methods which mimics a part of spin polarization in an uncontrolled manner akin to the partial success of UHF in describing dissociation. On the other hand, $\Delta$HF performs poorly for this problem suggesting that a proper estimation of the energetics lies in the state specific orbital optimization leading to spin-contamination in the presence of correlation. We opine that the K$_{if}$ value followed by or in conjunction with $\Delta$SCF methods with suggested XC functionals can be a computationally affordable preliminary screening method. This may be followed by TDDFT(D) (with recommended functionals), EOM-CCSD/CC3 and eventually FICMRCISD/EOM-CCSDT (if necessary) for the most promising candidates. Based on our studies in part I and part II of this work, we suggest the following screening pipeline: K$_{if}$ $\to$ $\Delta$SCF $\to$ TDDFT(D) $\to$ EOM-CCSD/CC3 $\to$ FICMRCISD. Evaluation of ST gaps also need to be supplemented by the vibronic coupling and rate constant computations before a decisive suggestion can be made alongwith consideration of other physical and chemical properties amendable to OLED applications.  

\section*{Supporting Information}
Cartesian geometries (XYZ structures) optimized at B3LYP/def2-TZVP level for all 9 molecules. Additionally, the complete numerical data of the signed errors in S$_1$-S$_0$, T$_1$-S$_0$ and $\Delta$E$_{ST}$ with respect to FICMRCI for 3 template molecules (azine-1N,4N and 7N) and signed errors with respect to EOM-CCSD for total set of 9 molecules to support the arguments in the paper. The K$_{if}$ inetgral along with the HOMO-LUMO gap for all molecules for a few suitable XC functionals are also given.

\section*{Acknowledgements}
SS gratefully acknowledges funding support from DST SERB, New Delhi, India (SRG/2021/000737) and IISER Kolkata. SC want to thank IISER Kolkata for Junior Research Fellowship. Prof. Ashwani K. Tiwari is acknowledged for sharing his computational facilities.  
\bibliographystyle{achemso}
\bibliography{invest.bib}


\end{document}